\documentclass[reprint, amsmath, amssymb, aps, prb]{revtex4-2}

\usepackage{graphicx}
\usepackage{dcolumn}
\usepackage[caption=false,font=small]{subfig}
\usepackage{bbold}

\makeatletter
\def\maketitle{
\@author@finish
\title@column\titleblock@produce
\suppressfloats[t]}
\makeatother

\def\cro{Ca$_{10}$Cr$_7$O$_{28}$}

\begin{document}

\title{Origin of the intermediate-temperature magnetic specific heat capacity in the spin-liquid candidate \cro}

\author{Joe Crossley}
 \email{jc413@st-andrews.ac.uk (corresponding author)}
 \affiliation{%
SUPA, School of Physics \& Astronomy, University of St Andrews, North Haugh, St Andrews, Fife KY16 9SS, U.K.}%
\author{Chris Hooley}
 \email{hooley@pks.mpg.de}
 \affiliation{%
Max Planck Institute for the Physics of Complex Systems, Nöthnitzer Straße 38, 01187 Dresden, Germany}%

\date{\today}

\begin{abstract}
We present several approximate calculations of the specific heat capacity of the model for \cro\ proposed by Balz \textit{et~al}.~[Phys.\,Rev.\,B \textbf{95}, 174414 (2017)], using methods including exact diagonalization, Thermal Pure Quantum States, and high-temperature expansions. In none of these cases are we able to reproduce the magnitude of the zero-field specific heat capacity shown in the intermediate-temperature $({\sim}\,5-15\,\rm{K})$ experimental data.  We discuss possible reasons for the discrepancy, and what it might tell us about the magnetic Hamiltonian for \cro.
\end{abstract}

\maketitle

\section{\label{sec:level3}Introduction}
This year marks the $50^{\rm{th}}$ anniversary of Anderson's famous suggestion \cite{anderson1973resonating} that the ground state of the spin-1/2 Heisenberg model on the triangular lattice might be a resonating valence bond state.  Despite the now convincing evidence that the ground state of that particular model has N{\'e}el order \cite{jolicoeur1989spin,bernu1994exact,capriotti1999long}, Anderson's proposal --- together with the discovery thirteen years later of the high-$T_{c}$ cuprate family of materials \cite{bednorz1986possible,wu1987superconductivity} --- provoked an interest in the study of so-called `spin liquid' phases which persists to this day \cite{savary2016quantum,broholm2020quantum}.

One candidate spin-liquid material that has given rise to considerable interest over the past few years is \cro. Balz~\textit{et~al}.~have suggested a two-dimensional model \cite{balz2016physical,balz2017crystal,balz2017magnetic,sonnenschein2019signatures}, while Alshalawi~\textit{et~al.}~recently proposed a 3D description \cite{alshalawi2022coexistence}. Both of these sets of authors attribute the magnetic behavior of \cro\ to the six $\mathrm{Cr}^{5+}$ ions in each formula unit \cite{balz2017crystal}. These are arranged in distorted kagome planes, with each site hosting a well-localized spin-1/2 degree of freedom.

\begin{figure}
\includegraphics[scale=0.3]{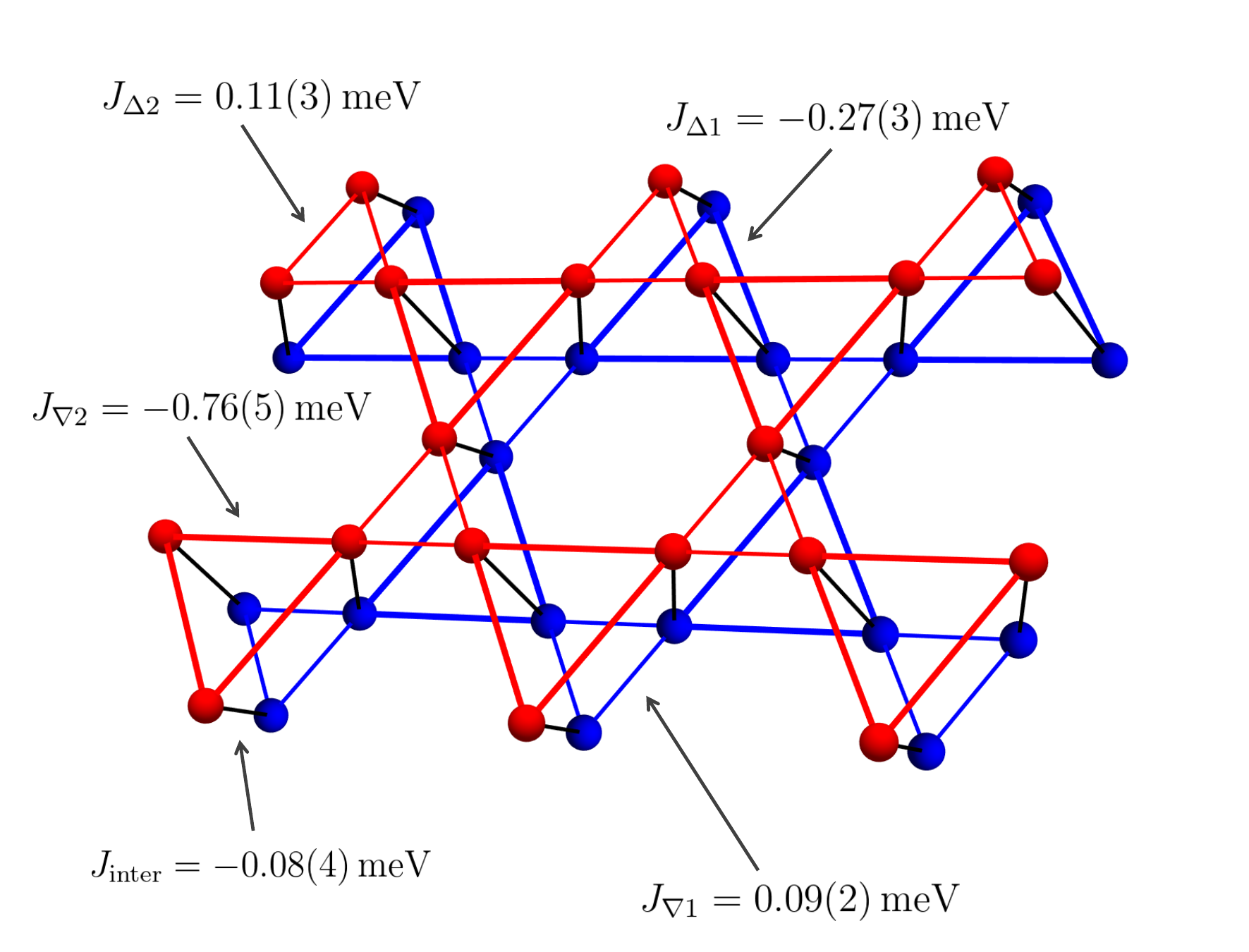}
\caption{\label{structure} A section from a single bilayer of \cro. The blue and red spheres denote $\mathrm{Cr}^{5+}$ ions from layers 1 and 2 respectively. All distinct couplings in the breathing bilayer kagome model are indicated by their symbol and associated value, as determined by Balz~\textit{et~al.}~in ref.\,\cite{balz2017magnetic}.}
\end{figure}

The model proposed by Balz~\textit{et~al.}~is known as the breathing bilayer kagome (BBK) Hamiltonian,
\begin{equation}
    \hat{H}=\sum_{\langle i, j\rangle}J_{ij}\hat{\bf{S}}_i\cdot\hat{\bf{S}}_j - g \mu_{B} h\sum_{i}{\hat{S}^z}_i,
\label{BBK}
\end{equation}
where $\langle i, j\rangle$ denotes summation over nearest neighbor bonds (counting each bond only once), $h$ denotes the applied magnetic field strength, and the $g$-factor is taken to be 2. This 2D model couples kagome layers into bilayers as shown in Fig.~\ref{structure}. The term `breathing' refers to the inequivalence of the differently oriented triangles within each kagome plane. The couplings themselves are of weak Heisenberg type, and have mixed ferro-/antiferromagnetic nature. Their numerical values, given by Balz~\textit{et~al.}, are included in Fig.~\ref{structure}; Balz~\textit{et~al.}~determined these by fitting the structure factor of the model to the spin-wave spectrum of \cro\ under a high magnetic field.

There is good reason to believe that both the model (BBK) and the corresponding material (\cro) exhibit spin liquid behavior at the lowest temperatures \cite{pohle2021theory, kshetrimayum2020tensor, biswas2018semiclassical,schmoll2022finite, balz2016physical, balz2017magnetic, balodhi2017synthesis, sonnenschein2019signatures}. The purpose of this paper isn't to question those beliefs, but rather to present evidence that the BBK model is appreciably incomplete as a description of the magnetism of \cro. We are not the first to raise concerns along these lines; Pohle~\textit{et~al.}~\cite{pohle2021theory} noted that the presence of finite-energy scattering weight at $\bf{q}=\bf{0}$ in zero-field neutron scattering experiments implies that there is significant anisotropy in the exchange interactions, a feature that the model (\ref{BBK}) does not reproduce. In this paper we further test this model by using various methods to calculate its specific heat capacity (with and without a $12\,\rm{T}$ field); we shall focus especially on the zero-field specific heat capacity because it is here that the inconsistency between the model and experiment is most clearly seen. Balz~\textit{et~al}.~also discuss, though ultimately discard, an alternative model called the `coupled hexagon model' \cite{balz2017crystal}. This model, too, appears unable to reproduce the specific heat capacity of \cro\ (results presented in Supplemental Material \cite{SM}).

The remainder of this paper is structured as follows. In Section II, we present the results of four different methods used to approximately calculate the specific heat capacity of the BBK model: the strong-triangle approximation (an approach based on the hierarchy of coupling strengths), exact diagonalization, Thermal Pure Quantum States, and the high-temperature expansion. In Section III we discuss our results, and the modifications to the magnetic model of \cro~that they might suggest.

\section{\label{sec:level3}Results}
\subsection{Strong-Triangle Approximation}
We expect something of a separation of scales in the BBK model since the $J_{\Delta 1}$ and $J_{\nabla 2}$ couplings stand out as stronger than the rest. This allows one to calculate the dominant contribution to the high-temperature specific heat capacity by switching off all other couplings, leaving isolated `strong triangles'. The unit cell of the bilayer lattice contains exactly one of each of the two types of strong triangle. These unit cells are non-interacting in this approximation and thus the specific heat capacity of the whole structure can be inferred from the partition function of a single unit cell,
\begin{equation}
    \mathcal{Z}_{\rm{cell}}(\beta)=\left(\sum_i e^{-\beta \epsilon^{\Delta 1}_i(h)}\right)\left(\sum_j e^{-\beta \epsilon^{\nabla 2}_j(h)}\right),
\label{Z}
\end{equation}
where $\epsilon^{\Delta 1}_i(h)$ and $\epsilon^{\nabla 2}_j(h)$ are the energy levels of each of the two types of triangle, calculated by diagonalizing the relevant $8\times8$ matrix obtained from (\ref{BBK}).

\begin{figure}[h]
\includegraphics[scale=1]{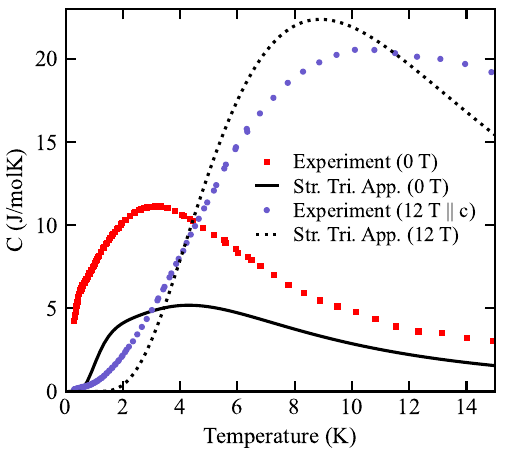}
\caption{\label{stm} The specific heat capacity of the breathing bilayer kagome model (1) using the strong-triangle approximation, in zero magnetic field (solid black curve) and in a magnetic field of $12\,\rm{T}$ (dotted black curve). Experimental data, previously published in ref.\,\cite{balz2017magnetic}, are also shown for comparison.}
\end{figure}

We show our results in Fig.~\ref{stm} for the $0\,\rm{T}$ and $12\,\rm{T}$ specific heat capacities calculated using this technique. Clearly the approximation works fairly well in the presence of a $12\,\rm{T}$ field. We expect this to be the case as the gap between eigenstates is well accounted for by the Zeeman and stronger ferromagnetic terms alone. In the absence of any external field, on the other hand, gaps between the lowest energy eigenstates occur only if we include the weaker bonds. Thus, our strong-triangle approximation cannot correctly reproduce the zero-field specific heat capacity at the lowest temperatures. This is no surprise; more concerningly, however, it shows significant disagreement with the experimental data even in the intermediate temperature range $(5-15\,\rm{K})$. Could this be due purely to the omission of the weaker exchange couplings, or must other physics be invoked to account for it?
\subsection{Exact Diagonalization}
\begin{figure}[h!]
\includegraphics[scale=1]{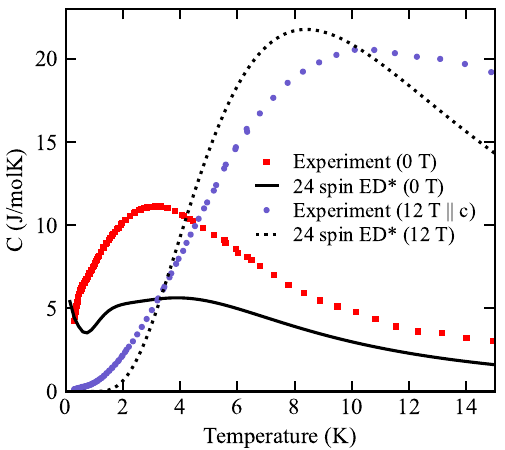}
\caption{\label{ed} The specific heat capacity of the breathing bilayer kagome model (1) calculated using exact diagonalization of a cluster of 24 spins (12 per layer), in zero magnetic field (solid black curve) and in a magnetic field of $12\,\rm{T}$ (dotted black curve). ``ED*" denotes that the exact diagonalization was performed with $J_{\mathrm{inter}}=0$. Experimental data, previously published in ref.\,\cite{balz2017magnetic}, are shown for comparison.}
\end{figure}
In an initial attempt to recover something of the weak coupling physics we now `switch on' $J_{\nabla 1}$ and $J_{\Delta 2}$, neglecting only $J_{\mathrm{inter}}$. This enables a simple exact diagonalization description where one diagonalizes a cluster of spins in each layer and then combines their spectra to calculate the specific heat capacity (see Supplemental
Material \cite{SM} for cluster dimensions). The calculation ends up looking much like that of the first technique:
\begin{equation}
    \mathcal{Z}_{\rm{cluster}}(\beta)=\left(\sum_i e^{-\beta \epsilon^{1}_i(h)}\right)\left(\sum_j e^{-\beta \epsilon^{2}_j(h)}\right),
\label{Z2}
\end{equation}
the difference being that the energy levels $\epsilon^{1}_i(h)$ and $\epsilon^{2}_j(h)$ are found by diagonalizing (\ref{BBK}) for not three but twelve spins in layers 1 and 2, respectively.

We show our results for a cluster of 24 spins in Fig.~\ref{ed}. Comparison with Fig.~\ref{stm} confirms that the inclusion of $J_{\nabla 1}$ and $J_{\Delta 2}$ causes no significant change to the $12\,\rm{T}$ curve. The story is somewhat similar for the zero-field curve, which appears to exactly follow that of the strong-triangle approximation down to ${\sim}\,4\,\rm{K}$, and still falls far short of the experimental specific heat capacity across almost the entire measured temperature range. We now turn to two more powerful techniques where all couplings are properly accounted for.
\subsection{Thermal Pure Quantum States}
The method of canonical Thermal Pure Quantum States (TPQS) was introduced by Sugiura and Shimizu \cite{sugiura2013canonical}. Fundamentally approximate, it can be used as an alternative to exact diagonalization, enabling larger cluster sizes to be treated. The method works by selecting a random vector $|\psi_0\rangle$ in the many-body Hilbert space and using this pure state as a kind of proxy for the true infinite-temperature mixed state. This `thermal state' is cooled down step-by-step like so:
\begin{equation}
    |k\rangle = (l-\hat{h}) |k-1\rangle =  (l-\hat{h})^k |\psi_0\rangle,
\label{cooldown}
\end{equation}
where $\hat{h} = \hat{H}/N$, $N$ is the number of lattice sites, and $l$ is an upper bound on the eigenvalue spectrum of $\hat{h}$. By taking expectation values at each iteration, thermodynamic variables can be calculated as functions of temperature. In our case, we found the heat capacity from the numerical temperature derivative of the energy per particle,
\begin{equation}
    \langle\hat{h}\rangle_{\beta,N} \approx \sum_{k=0}^{\infty} (N\beta)^{2k} \left(\frac{\langle k|\hat{h}|k\rangle}{(2k)!} + N\beta\frac{\langle k|\hat{h}|k+1\rangle}{(2k+1)!}\right),
\label{tpqs_energy}
\end{equation}
which we evaluated to sufficiently high order that the associated truncation error may reasonably be ignored compared to the uncertainty due to the initial starting state. Sugiura and Shimizu provide a method for estimating this latter uncertainty. We followed their scheme and ensured that the uncertainty margin on all our results never exceeded ${\sim}\,1$\%. We also bench-marked our code against the data of Elstner and Young for the spin-1/2 Heisenberg antiferromagnet on the kagome lattice \cite{elstner1994spin} (see Supplemental
Material \cite{SM}).

Fig.~\ref{tpqs} shows the zero-field TPQS specific heat capacity of (\ref{BBK}) for clusters of 18, 24, and 30 spins (see Supplemental Material \cite{SM} for exact cluster dimensions). The TPQS method becomes more accurate for larger Hilbert spaces. In the 30-spin case a single TPQS run was sufficient but for smaller clusters we averaged our results over several runs to bring our uncertainty measures down: eight runs for the 24-spin case and 100 runs for the 18-spin case. Our results were obtained by parallelizing the calculation across 64 cores of one of the nodes in the Edinburgh SGE cluster. Due to memory considerations, we successively generated the matrix elements of the Hamiltonian rather than storing them all simultaneously.

For spin clusters of these sizes, the separation of coupling strengths in the model (\ref{BBK}) is clearly manifest in the two distinct peaks in the specific heat capacity. No such double-peak structure is seen in the experimental data.
\begin{figure}[t!]
\includegraphics[scale=1]{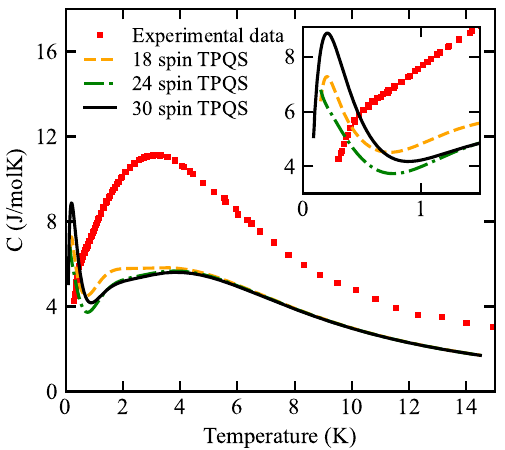}
\caption{\label{tpqs} The specific heat capacity of the breathing bilayer kagome model (1) calculated using Thermal Pure Quantum States for clusters of 18, 24, and 30 spins, in the absence of a magnetic field. For 18 spins we averaged across 100 runs, for 24 spins we averaged across eight runs, and for 30 spins only a single run was used. In all cases the uncertainty margin at each temperature never exceeds ${\sim}\,1$\%. The experimental zero-field specific heat capacity of \cro, previously published in ref.\,\cite{balz2017magnetic}, is also shown. For clarity, we have included an inset which enlarges the heat capacity curves at the lowest temperatures.}
\end{figure}
\nocite{schmidt2011eighth} 
\begin{figure*}[t!]
\centerline{\includegraphics[scale=1,trim={1.4cm 0 0 0},clip]{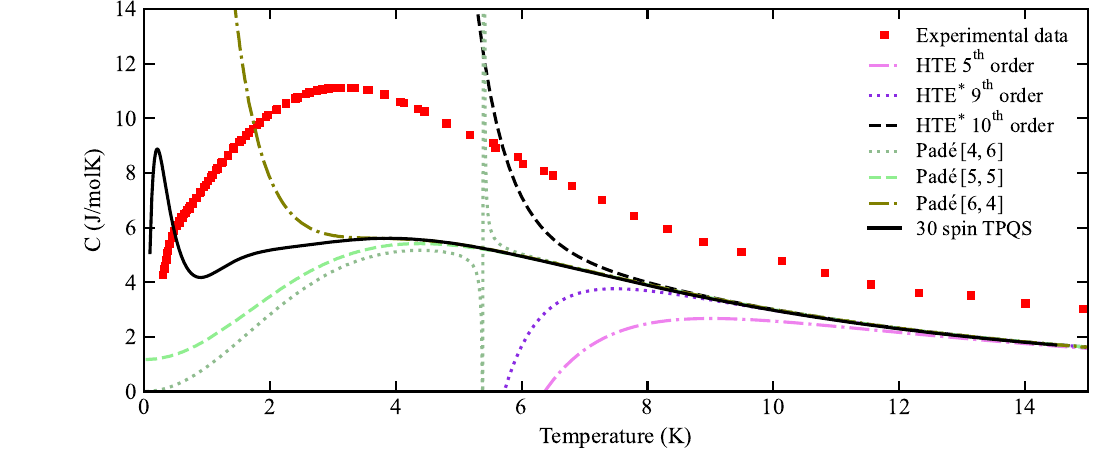}}
\caption{\label{hte} The specific heat capacity of the breathing bilayer kagome model (1) in the absence of a magnetic field, calculated using the high-temperature expansion to various orders as stated in the legend. The curves labelled ``HTE*'' were calculated using a software package \cite{lohmann2014tenth} allowing only four different coupling strengths, so we set $J_{\nabla1} = J_{\Delta2}$ in these cases (which is in any case consistent with the quoted uncertainties in these coupling constants). The various Pad{\'e} curves were also computed using the aforementioned software package; they are based on the $10^{\rm th}$ order expansion obtained by the software, labelled here as ``HTE* $10^{\rm th}$ order''. For comparison, we show our 30-spin Thermal Pure Quantum States curve for the model, as well as the experimental zero-field specific heat capacity data previously published in ref.\,\cite{balz2017magnetic}.}
\end{figure*}
\subsection{High-Temperature Expansion}
The high-temperature expansion (HTE) expresses the partition function as
\begin{equation}
    \mathcal{Z}(\beta) = \rm{Tr}[\mathbb{1}] -\beta \rm{Tr}[\hat{\it{H}}] + \frac{\beta^2}{2} \rm{Tr}[\hat{\it{H}}^2]- \frac{\beta^3}{3!} \rm{Tr}[\hat{\it{H}}^3]+\dots .
\end{equation}
For spin systems, the expansion coefficients decompose into much smaller traces over products of spin-spin bonds appearing in the Hamiltonian. However, these `moments' become increasingly complex for higher orders.

Following ref.\,\cite{schmidt2011eighth}, we calculated the first five moments for the heat capacity of (\ref{BBK}) by hand. The C++ package by Lohmann~\textit{et~al.}~\cite{lohmann2014tenth}~enabled us to further calculate the series up to tenth order, as well as providing three different Pad{\'e} approximants to extend the series. The package only allows for four different coupling strengths, so we set $J_{\nabla1} = J_{\Delta2}$ (which is in any case consistent with the quoted uncertainties in these coupling constants) in these cases.

We show our results in Fig.~\ref{hte} along with the experimental zero-field specific heat capacity and the 30-spin TPQS curve for comparison. Our finite-order results diverge at low temperatures as is normal for such series. Assuming further orders in the expansion are well behaved, one typically decides the temperature down to which such curves can be trusted by looking for the point at which consecutive orders separate appreciably from one another. The geometry of the BBK lattice gives no specific reason to expect the higher-order terms to be poorly behaved, so we take the ninth- or tenth-order curves as essentially correct down to ${\sim}\,9\,{\rm K}$.

The three Pad{\'e} approximants agree down to even lower temperatures. These approximants are quotients of polynomials in $J/(k_{B}T)$ and we label them Pad{\'e}\,$[L,M]$, where $L$ is the polynomial order of the numerator and $M$ is the polynomial order of the denominator. A `complete' approximation to the heat capacity should converge to zero as $T\rightarrow0$; this is only guaranteed for approximants where $L<M$\ \cite{elstner1994spin}. Yet, our curve for which this condition is satisfied (Pad{\'e}\,$[4,6]$) exhibits a dramatic singularity at intermediate temperatures. Comparison with the other approximants, as well as with the TPQS curve, suggests that this is a ``spurious pole'' (a well-known effect where the Pad{\'e} approximant exhibits a pole that is not a feature of the target function). In any case, we find the three approximants in good agreement with one another right down to $\sim\,6\,\rm{K}$.

\section{\label{sec:level3}Discussion}
All this has little to say to those interested in the spin liquid phase of the BBK model; no method used in this study is expected to work well at such low temperatures. However, our calculations have clear implications for those who would use the model to describe the spin liquid phase of \cro: if the model doesn't describe the material's magnetism well at these intermediate temperatures why should we expect it to perform well at the lowest temperatures?

The strong-triangle approximation is clearly a drastic one, especially in the absence of a magnetic field. Even so, it is very useful: the one who believes that \cro\ is described by the BBK model (\ref{BBK}) is forced to say that the weak couplings would, were they included, account for the difference between the calculated and measured specific heat capacities of Fig.\,\ref{stm}. This requires that the weak couplings strongly enhance the specific heat capacity far beyond their temperature scale of ${\sim}\,1\,\rm{K}$, which intuitively seems unlikely given the structure of the model. To test this intuition we undertook three further calculations, all of which sought to include the effect of the weaker couplings.

Our exact diagonalization method includes all but one of the weaker coupling types, and the TPQS method includes them all. As expected, these weaker couplings lead to no noticeable change in the specific heat capacity right down to ${\sim}\,4\,\rm{K}$; rather, a second peak emerges below $1\,\rm{K}$, increasing the apparent disparity between theory and experiment. Yet both methods are subject to finite-size effects. We might question whether longer-range physics would push this new low-temperature peak upward to supplement the broader peak at ${\sim}\,4\,\rm{K}$ --- from an entropic perspective, this peak-migration effect could account for the difference between theory and experiment (see Supplemental Material \cite{SM}).

To address the issue of potential finite-size effects we now turn to our final technique, the high-temperature expansion, which allows us to calculate the specific heat capacity in the thermodynamic limit. The ninth- and tenth-order HTEs follow our 30-spin TPQS curve down to  ${\sim}\,9\,\rm{K}$ (below which we argue they are no longer valid anyway), and all the Pad{\'e} extensions follow our 30-spin TPQS curve down to ${\sim}\,6\,\rm{K}$. Thus, despite finite-size effects, we claim that the 30-spin TPQS curve suitably reproduces the specific heat capacity of the BBK model (\ref{BBK}) in the thermodynamic limit, right down to ${\sim}\,6\,\rm{K}$.

We are left with an apparent disparity between theory and experiment; what are we to make of it? One might seek a resolution by claiming that there remains a residual phononic contribution in the experimental data. The size of the discrepancy, and the fact that it appears to be strongly dependent on the applied magnetic field, make it unlikely that this is the full story.

Another possibility is that the BBK model (\ref{BBK}) is correct in spirit, but that the exact values of the five different isotropic couplings require further optimization in order to reproduce not only the 11 T neutron scattering data of \cro\ but also the specific heat capacity data. To explore this possibility fully would require a thorough variational search across the five-dimensional parameter space of the BBK model. Such an investigation is beyond the scope of this paper. However, as a test we can attempt to reverse the order of the procedure for determining the coupling constants. Balz~\textit{et~al}.~fitted the coupling constants to the single-spin-flip dispersion relations obtained via high-field inelastic neutron scattering; as we have shown, the resulting coupling constants do not account for the zero-field specific heat capacity.  We have instead determined several sets of possible coupling constants via a fit to the high-temperature tail of the 0 T specific heat capacity, but in none of these cases do we find a reasonable fit to the single-spin-flip dispersion relations (see Supplemental Material \cite{SM}). Taking these results together with Pohle~\textit{et~al}.'s\ argument for anisotropy in the exchange interactions (finite-energy scattering weight at $\bf{q}=\bf{0}$)\ \cite{pohle2021theory}, we tentatively conclude that there is unlikely to be a pure Heisenberg model that reproduces all of the currently available experimental data on \cro.

A weak Dzyaloshinskii-Moriya interaction \cite{blundell2001magnetism, pohle2021theory, schmoll2022finite} would break the degeneracy of the ground state manifold of each of the strong triangles, leaving only a Kramers doublet in its place. This could be the mechanism by which the entropy under the low temperature peak is moved to supplement the broader peak at higher temperatures. While such anisotropic interactions are not usually essential to describe the magnetism of 3d transition metal ions, the quoted strengths of the Heisenberg interactions in the BBK model are themselves so weak that lower symmetry effects need not be that great before they become important. One might oppose the suggestion of anisotropy by pointing to the similarity of the magnetic susceptibility curves of Balz~{\it et~al.\/}~\cite{balz2017magnetic} when the field is applied along different crystal axis directions. However, local anisotropic effects in competition can conspire to produce exactly such behavior \cite{inprep1}.

Alongside microscopic calculations based on a tight-binding expansion, we are currently exploring a variational approach based on the addition of general anisotropic exchange terms to the strong-triangle approximation, fitting to the wealth of experimental data available. While our plan is initially to focus on the strong bonds in the model, we hope this will also function to improve the accuracy of the model down to the lowest temperatures \cite{inprep1}.
\vspace{-4mm}
\begin{acknowledgments}
The authors would like to thank Christian Balz, Bella Lake, and Nic Shannon for helpful discussions, and Christian Balz and Bella Lake for kindly sharing the original specific heat capacity and inelastic neutron scattering data published in ref.~\cite{balz2017magnetic}. We also thank A. Lohmann, J. Richter,  and H.-J. Schmidt for their excellent C++ package for computing high-temperature expansions \cite{lohmann2014tenth}, which can be found at http://www.uni-magdeburg.de/jschulen/HTE10/.~All Thermal Pure Quantum States calculations were performed on the Edinburgh University SGE cluster.

JAC acknowledges support from UKRI via EPSRC grant EP/T518062/1. CAH acknowledges support from UKRI via EPSRC grant EP/R031924/1.

This work was performed in part at Aspen Center for Physics, which is supported by National Science Foundation grant PHY-2210452.
\end{acknowledgments}
\bibliography{origin_cp_cro}

\begin{thebibliography}{27}%
\makeatletter
\providecommand \@ifxundefined [1]{%
 \@ifx{#1\undefined}
}%
\providecommand \@ifnum [1]{%
 \ifnum #1\expandafter \@firstoftwo
 \else \expandafter \@secondoftwo
 \fi
}%
\providecommand \@ifx [1]{%
 \ifx #1\expandafter \@firstoftwo
 \else \expandafter \@secondoftwo
 \fi
}%
\providecommand \natexlab [1]{#1}%
\providecommand \enquote  [1]{``#1''}%
\providecommand \bibnamefont  [1]{#1}%
\providecommand \bibfnamefont [1]{#1}%
\providecommand \citenamefont [1]{#1}%
\providecommand \href@noop [0]{\@secondoftwo}%
\providecommand \href [0]{\begingroup \@sanitize@url \@href}%
\providecommand \@href[1]{\@@startlink{#1}\@@href}%
\providecommand \@@href[1]{\endgroup#1\@@endlink}%
\providecommand \@sanitize@url [0]{\catcode `\\12\catcode `\$12\catcode
  `\&12\catcode `\#12\catcode `\^12\catcode `\_12\catcode `\%12\relax}%
\providecommand \@@startlink[1]{}%
\providecommand \@@endlink[0]{}%
\providecommand \url  [0]{\begingroup\@sanitize@url \@url }%
\providecommand \@url [1]{\endgroup\@href {#1}{\urlprefix }}%
\providecommand \urlprefix  [0]{URL }%
\providecommand \Eprint [0]{\href }%
\providecommand \doibase [0]{https://doi.org/}%
\providecommand \selectlanguage [0]{\@gobble}%
\providecommand \bibinfo  [0]{\@secondoftwo}%
\providecommand \bibfield  [0]{\@secondoftwo}%
\providecommand \translation [1]{[#1]}%
\providecommand \BibitemOpen [0]{}%
\providecommand \bibitemStop [0]{}%
\providecommand \bibitemNoStop [0]{.\EOS\space}%
\providecommand \EOS [0]{\spacefactor3000\relax}%
\providecommand \BibitemShut  [1]{\csname bibitem#1\endcsname}%
\let\auto@bib@innerbib\@empty
\bibitem [{\citenamefont {Anderson}(1973)}]{anderson1973resonating}%
  \BibitemOpen
  \bibfield  {author} {\bibinfo {author} {\bibfnamefont {P.~W.}\ \bibnamefont
  {Anderson}},\ }\bibfield  {title} {\bibinfo {title} {Resonating valence
  bonds: A new kind of insulator?},\ }\href@noop {} {\bibfield  {journal}
  {\bibinfo  {journal} {Mater. Res. Bull.}\ }\textbf {\bibinfo {volume} {8}},\
  \bibinfo {pages} {153} (\bibinfo {year} {1973})}\BibitemShut {NoStop}%
\bibitem [{\citenamefont {Jolicoeur}\ and\ \citenamefont
  {Le~Guillou}(1989)}]{jolicoeur1989spin}%
  \BibitemOpen
  \bibfield  {author} {\bibinfo {author} {\bibfnamefont {T.}~\bibnamefont
  {Jolicoeur}}\ and\ \bibinfo {author} {\bibfnamefont {J.~C.}\ \bibnamefont
  {Le~Guillou}},\ }\bibfield  {title} {\bibinfo {title} {Spin-wave results for
  the triangular {H}eisenberg antiferromagnet},\ }\href@noop {} {\bibfield
  {journal} {\bibinfo  {journal} {Phys. Rev. B}\ }\textbf {\bibinfo {volume}
  {40}},\ \bibinfo {pages} {2727} (\bibinfo {year} {1989})}\BibitemShut
  {NoStop}%
\bibitem [{\citenamefont {Bernu}\ \emph {et~al.}(1994)\citenamefont {Bernu},
  \citenamefont {Lecheminant}, \citenamefont {Lhuillier},\ and\ \citenamefont
  {Pierre}}]{bernu1994exact}%
  \BibitemOpen
  \bibfield  {author} {\bibinfo {author} {\bibfnamefont {B.}~\bibnamefont
  {Bernu}}, \bibinfo {author} {\bibfnamefont {P.}~\bibnamefont {Lecheminant}},
  \bibinfo {author} {\bibfnamefont {C.}~\bibnamefont {Lhuillier}},\ and\
  \bibinfo {author} {\bibfnamefont {L.}~\bibnamefont {Pierre}},\ }\bibfield
  {title} {\bibinfo {title} {Exact spectra, spin susceptibilities, and order
  parameter of the quantum {H}eisenberg antiferromagnet on the triangular
  lattice},\ }\href@noop {} {\bibfield  {journal} {\bibinfo  {journal} {Phys.
  Rev. B}\ }\textbf {\bibinfo {volume} {50}},\ \bibinfo {pages} {10048}
  (\bibinfo {year} {1994})}\BibitemShut {NoStop}%
\bibitem [{\citenamefont {Capriotti}\ \emph {et~al.}(1999)\citenamefont
  {Capriotti}, \citenamefont {Trumper},\ and\ \citenamefont
  {Sorella}}]{capriotti1999long}%
  \BibitemOpen
  \bibfield  {author} {\bibinfo {author} {\bibfnamefont {L.}~\bibnamefont
  {Capriotti}}, \bibinfo {author} {\bibfnamefont {A.~E.}\ \bibnamefont
  {Trumper}},\ and\ \bibinfo {author} {\bibfnamefont {S.}~\bibnamefont
  {Sorella}},\ }\bibfield  {title} {\bibinfo {title} {Long-range {N}{\'e}el
  {O}rder in the {T}riangular {H}eisenberg {M}odel},\ }\href@noop {} {\bibfield
   {journal} {\bibinfo  {journal} {Phys. Rev. Lett.}\ }\textbf {\bibinfo
  {volume} {82}},\ \bibinfo {pages} {3899} (\bibinfo {year}
  {1999})}\BibitemShut {NoStop}%
\bibitem [{\citenamefont {Bednorz}\ and\ \citenamefont
  {M{\"u}ller}(1986)}]{bednorz1986possible}%
  \BibitemOpen
  \bibfield  {author} {\bibinfo {author} {\bibfnamefont {J.~G.}\ \bibnamefont
  {Bednorz}}\ and\ \bibinfo {author} {\bibfnamefont {K.~A.}\ \bibnamefont
  {M{\"u}ller}},\ }\bibfield  {title} {\bibinfo {title} {Possible high {$T_c$}
  superconductivity in the {Ba-La-Cu-O} system},\ }\href@noop {} {\bibfield
  {journal} {\bibinfo  {journal} {Z. Phys. B}\ }\textbf {\bibinfo {volume}
  {64}},\ \bibinfo {pages} {189} (\bibinfo {year} {1986})}\BibitemShut
  {NoStop}%
\bibitem [{\citenamefont {Wu}\ \emph {et~al.}(1987)\citenamefont {Wu},
  \citenamefont {Ashburn}, \citenamefont {Torng}, \citenamefont {Hor},
  \citenamefont {Meng}, \citenamefont {Gao}, \citenamefont {Huang},
  \citenamefont {Wang},\ and\ \citenamefont {Chu}}]{wu1987superconductivity}%
  \BibitemOpen
  \bibfield  {author} {\bibinfo {author} {\bibfnamefont {M.-K.}\ \bibnamefont
  {Wu}}, \bibinfo {author} {\bibfnamefont {J.~R.}\ \bibnamefont {Ashburn}},
  \bibinfo {author} {\bibfnamefont {C.}~\bibnamefont {Torng}}, \bibinfo
  {author} {\bibfnamefont {P.-H.}\ \bibnamefont {Hor}}, \bibinfo {author}
  {\bibfnamefont {R.~L.}\ \bibnamefont {Meng}}, \bibinfo {author}
  {\bibfnamefont {L.}~\bibnamefont {Gao}}, \bibinfo {author} {\bibfnamefont
  {Z.~J.}\ \bibnamefont {Huang}}, \bibinfo {author} {\bibfnamefont
  {Y.}~\bibnamefont {Wang}},\ and\ \bibinfo {author} {\bibfnamefont
  {a.}~\bibnamefont {Chu}},\ }\bibfield  {title} {\bibinfo {title}
  {Superconductivity at 93 {K} in a new mixed-phase {Y-Ba-Cu-O} compound system
  at ambient pressure},\ }\href@noop {} {\bibfield  {journal} {\bibinfo
  {journal} {Phys. Rev. Lett.}\ }\textbf {\bibinfo {volume} {58}},\ \bibinfo
  {pages} {908} (\bibinfo {year} {1987})}\BibitemShut {NoStop}%
\bibitem [{\citenamefont {Savary}\ and\ \citenamefont
  {Balents}(2016)}]{savary2016quantum}%
  \BibitemOpen
  \bibfield  {author} {\bibinfo {author} {\bibfnamefont {L.}~\bibnamefont
  {Savary}}\ and\ \bibinfo {author} {\bibfnamefont {L.}~\bibnamefont
  {Balents}},\ }\bibfield  {title} {\bibinfo {title} {Quantum spin liquids: a
  review},\ }\href@noop {} {\bibfield  {journal} {\bibinfo  {journal} {Rep.
  Prog. Phys.}\ }\textbf {\bibinfo {volume} {80}},\ \bibinfo {pages} {016502}
  (\bibinfo {year} {2016})}\BibitemShut {NoStop}%
\bibitem [{\citenamefont {Broholm}\ \emph {et~al.}(2020)\citenamefont
  {Broholm}, \citenamefont {Cava}, \citenamefont {Kivelson}, \citenamefont
  {Nocera}, \citenamefont {Norman},\ and\ \citenamefont
  {Senthil}}]{broholm2020quantum}%
  \BibitemOpen
  \bibfield  {author} {\bibinfo {author} {\bibfnamefont {C.}~\bibnamefont
  {Broholm}}, \bibinfo {author} {\bibfnamefont {R.}~\bibnamefont {Cava}},
  \bibinfo {author} {\bibfnamefont {S.}~\bibnamefont {Kivelson}}, \bibinfo
  {author} {\bibfnamefont {D.}~\bibnamefont {Nocera}}, \bibinfo {author}
  {\bibfnamefont {M.}~\bibnamefont {Norman}},\ and\ \bibinfo {author}
  {\bibfnamefont {T.}~\bibnamefont {Senthil}},\ }\bibfield  {title} {\bibinfo
  {title} {Quantum spin liquids},\ }\href@noop {} {\bibfield  {journal}
  {\bibinfo  {journal} {Science}\ }\textbf {\bibinfo {volume} {367}},\ \bibinfo
  {pages} {eaay0668} (\bibinfo {year} {2020})}\BibitemShut {NoStop}%
\bibitem [{\citenamefont {Balz}\ \emph {et~al.}(2016)\citenamefont {Balz} \emph
  {et~al.}}]{balz2016physical}%
  \BibitemOpen
  \bibfield  {author} {\bibinfo {author} {\bibfnamefont {C.}~\bibnamefont
  {Balz}} \emph {et~al.},\ }\bibfield  {title} {\bibinfo {title} {Physical
  realization of a quantum spin liquid based on a complex frustration
  mechanism},\ }\href@noop {} {\bibfield  {journal} {\bibinfo  {journal}
  {Nature Physics}\ }\textbf {\bibinfo {volume} {12}},\ \bibinfo {pages} {942}
  (\bibinfo {year} {2016})}\BibitemShut {NoStop}%
\bibitem [{\citenamefont {Balz}\ \emph
  {et~al.}(2017{\natexlab{a}})\citenamefont {Balz}, \citenamefont {Lake},
  \citenamefont {Reehuis}, \citenamefont {Nazmul~Islam}, \citenamefont
  {Prokhnenko}, \citenamefont {Singh}, \citenamefont {Pattison},\ and\
  \citenamefont {T{\'o}th}}]{balz2017crystal}%
  \BibitemOpen
  \bibfield  {author} {\bibinfo {author} {\bibfnamefont {C.}~\bibnamefont
  {Balz}}, \bibinfo {author} {\bibfnamefont {B.}~\bibnamefont {Lake}}, \bibinfo
  {author} {\bibfnamefont {M.}~\bibnamefont {Reehuis}}, \bibinfo {author}
  {\bibfnamefont {A.~T.~M.}\ \bibnamefont {Nazmul~Islam}}, \bibinfo {author}
  {\bibfnamefont {O.}~\bibnamefont {Prokhnenko}}, \bibinfo {author}
  {\bibfnamefont {Y.}~\bibnamefont {Singh}}, \bibinfo {author} {\bibfnamefont
  {P.}~\bibnamefont {Pattison}},\ and\ \bibinfo {author} {\bibfnamefont
  {S.}~\bibnamefont {T{\'o}th}},\ }\bibfield  {title} {\bibinfo {title}
  {Crystal growth, structure and magnetic properties of \cro},\ }\href@noop {}
  {\bibfield  {journal} {\bibinfo  {journal} {J. Phys. Condens. Matter}\
  }\textbf {\bibinfo {volume} {29}},\ \bibinfo {pages} {225802} (\bibinfo
  {year} {2017}{\natexlab{a}})}\BibitemShut {NoStop}%
\bibitem [{\citenamefont {Balz}\ \emph
  {et~al.}(2017{\natexlab{b}})\citenamefont {Balz}, \citenamefont {Lake},
  \citenamefont {Nazmul~Islam}, \citenamefont {Singh}, \citenamefont
  {Rodriguez-Rivera}, \citenamefont {Guidi}, \citenamefont {Wheeler},
  \citenamefont {Simeoni},\ and\ \citenamefont {Ryll}}]{balz2017magnetic}%
  \BibitemOpen
  \bibfield  {author} {\bibinfo {author} {\bibfnamefont {C.}~\bibnamefont
  {Balz}}, \bibinfo {author} {\bibfnamefont {B.}~\bibnamefont {Lake}}, \bibinfo
  {author} {\bibfnamefont {A.~T.~M.}\ \bibnamefont {Nazmul~Islam}}, \bibinfo
  {author} {\bibfnamefont {Y.}~\bibnamefont {Singh}}, \bibinfo {author}
  {\bibfnamefont {J.~A.}\ \bibnamefont {Rodriguez-Rivera}}, \bibinfo {author}
  {\bibfnamefont {T.}~\bibnamefont {Guidi}}, \bibinfo {author} {\bibfnamefont
  {E.~M.}\ \bibnamefont {Wheeler}}, \bibinfo {author} {\bibfnamefont {G.~G.}\
  \bibnamefont {Simeoni}},\ and\ \bibinfo {author} {\bibfnamefont
  {H.}~\bibnamefont {Ryll}},\ }\bibfield  {title} {\bibinfo {title} {Magnetic
  {H}amiltonian and phase diagram of the quantum spin liquid \cro},\
  }\href@noop {} {\bibfield  {journal} {\bibinfo  {journal} {Phys. Rev. B}\
  }\textbf {\bibinfo {volume} {95}},\ \bibinfo {pages} {174414} (\bibinfo
  {year} {2017}{\natexlab{b}})}\BibitemShut {NoStop}%
\bibitem [{\citenamefont {Sonnenschein}\ \emph {et~al.}(2019)\citenamefont
  {Sonnenschein}, \citenamefont {Balz}, \citenamefont {Tutsch}, \citenamefont
  {Lang}, \citenamefont {Ryll}, \citenamefont {Rodriguez-Rivera}, \citenamefont
  {{A. T. M. Nazmul Islam}}, \citenamefont {Lake},\ and\ \citenamefont
  {Reuther}}]{sonnenschein2019signatures}%
  \BibitemOpen
  \bibfield  {author} {\bibinfo {author} {\bibfnamefont {J.}~\bibnamefont
  {Sonnenschein}}, \bibinfo {author} {\bibfnamefont {C.}~\bibnamefont {Balz}},
  \bibinfo {author} {\bibfnamefont {U.}~\bibnamefont {Tutsch}}, \bibinfo
  {author} {\bibfnamefont {M.}~\bibnamefont {Lang}}, \bibinfo {author}
  {\bibfnamefont {H.}~\bibnamefont {Ryll}}, \bibinfo {author} {\bibfnamefont
  {J.~A.}\ \bibnamefont {Rodriguez-Rivera}}, \bibinfo {author} {\bibnamefont
  {{A. T. M. Nazmul Islam}}}, \bibinfo {author} {\bibfnamefont
  {B.}~\bibnamefont {Lake}},\ and\ \bibinfo {author} {\bibfnamefont
  {J.}~\bibnamefont {Reuther}},\ }\bibfield  {title} {\bibinfo {title}
  {Signatures for spinons in the quantum spin liquid candidate \cro},\
  }\href@noop {} {\bibfield  {journal} {\bibinfo  {journal} {Phys. Rev. B}\
  }\textbf {\bibinfo {volume} {100}},\ \bibinfo {pages} {174428} (\bibinfo
  {year} {2019})}\BibitemShut {NoStop}%
\bibitem [{\citenamefont {Alshalawi}\ \emph {et~al.}(2022)\citenamefont
  {Alshalawi}, \citenamefont {Alonso}, \citenamefont {Landa-C{\'a}novas},\ and\
  \citenamefont {de~la Presa}}]{alshalawi2022coexistence}%
  \BibitemOpen
  \bibfield  {author} {\bibinfo {author} {\bibfnamefont {D.~R.}\ \bibnamefont
  {Alshalawi}}, \bibinfo {author} {\bibfnamefont {J.~M.}\ \bibnamefont
  {Alonso}}, \bibinfo {author} {\bibfnamefont {A.~R.}\ \bibnamefont
  {Landa-C{\'a}novas}},\ and\ \bibinfo {author} {\bibfnamefont
  {P.}~\bibnamefont {de~la Presa}},\ }\bibfield  {title} {\bibinfo {title}
  {Coexistence of {T}wo {S}pin {F}rustration {P}athways in the {Q}uantum {S}pin
  {L}iquid \cro},\ }\href@noop {} {\bibfield  {journal} {\bibinfo  {journal}
  {Inorg. Chem.}\ }\textbf {\bibinfo {volume} {61}},\ \bibinfo {pages} {16228}
  (\bibinfo {year} {2022})}\BibitemShut {NoStop}%
\bibitem [{\citenamefont {Pohle}\ \emph {et~al.}(2021)\citenamefont {Pohle},
  \citenamefont {Yan},\ and\ \citenamefont {Shannon}}]{pohle2021theory}%
  \BibitemOpen
  \bibfield  {author} {\bibinfo {author} {\bibfnamefont {R.}~\bibnamefont
  {Pohle}}, \bibinfo {author} {\bibfnamefont {H.}~\bibnamefont {Yan}},\ and\
  \bibinfo {author} {\bibfnamefont {N.}~\bibnamefont {Shannon}},\ }\bibfield
  {title} {\bibinfo {title} {Theory of \cro~as a bilayer breathing-kagome
  magnet: {C}lassical thermodynamics and semiclassical dynamics},\ }\href@noop
  {} {\bibfield  {journal} {\bibinfo  {journal} {Phys. Rev. B}\ }\textbf
  {\bibinfo {volume} {104}},\ \bibinfo {pages} {024426} (\bibinfo {year}
  {2021})}\BibitemShut {NoStop}%
\bibitem [{\citenamefont {Kshetrimayum}\ \emph {et~al.}(2020)\citenamefont
  {Kshetrimayum}, \citenamefont {Balz}, \citenamefont {Lake},\ and\
  \citenamefont {Eisert}}]{kshetrimayum2020tensor}%
  \BibitemOpen
  \bibfield  {author} {\bibinfo {author} {\bibfnamefont {A.}~\bibnamefont
  {Kshetrimayum}}, \bibinfo {author} {\bibfnamefont {C.}~\bibnamefont {Balz}},
  \bibinfo {author} {\bibfnamefont {B.}~\bibnamefont {Lake}},\ and\ \bibinfo
  {author} {\bibfnamefont {J.}~\bibnamefont {Eisert}},\ }\bibfield  {title}
  {\bibinfo {title} {Tensor network investigation of the double layer {K}agome
  compound \cro},\ }\href@noop {} {\bibfield  {journal} {\bibinfo  {journal}
  {Ann. Phys.}\ }\textbf {\bibinfo {volume} {421}},\ \bibinfo {pages} {168292}
  (\bibinfo {year} {2020})}\BibitemShut {NoStop}%
\bibitem [{\citenamefont {Biswas}\ and\ \citenamefont
  {Damle}(2018)}]{biswas2018semiclassical}%
  \BibitemOpen
  \bibfield  {author} {\bibinfo {author} {\bibfnamefont {S.}~\bibnamefont
  {Biswas}}\ and\ \bibinfo {author} {\bibfnamefont {K.}~\bibnamefont {Damle}},\
  }\bibfield  {title} {\bibinfo {title} {Semiclassical theory for liquidlike
  behavior of the frustrated magnet \cro},\ }\href@noop {} {\bibfield
  {journal} {\bibinfo  {journal} {Phys. Rev. B}\ }\textbf {\bibinfo {volume}
  {97}},\ \bibinfo {pages} {115102} (\bibinfo {year} {2018})}\BibitemShut
  {NoStop}%
\bibitem [{\citenamefont {Schmoll}\ \emph {et~al.}(2022)\citenamefont
  {Schmoll}, \citenamefont {Balz}, \citenamefont {Lake}, \citenamefont
  {Eisert},\ and\ \citenamefont {Kshetrimayum}}]{schmoll2022finite}%
  \BibitemOpen
  \bibfield  {author} {\bibinfo {author} {\bibfnamefont {P.}~\bibnamefont
  {Schmoll}}, \bibinfo {author} {\bibfnamefont {C.}~\bibnamefont {Balz}},
  \bibinfo {author} {\bibfnamefont {B.}~\bibnamefont {Lake}}, \bibinfo {author}
  {\bibfnamefont {J.}~\bibnamefont {Eisert}},\ and\ \bibinfo {author}
  {\bibfnamefont {A.}~\bibnamefont {Kshetrimayum}},\ }\href@noop {} {\bibinfo
  {title} {Finite temperature tensor network algorithm for frustrated
  two-dimensional quantum materials}} (\bibinfo {year} {2022}),\ \Eprint
  {https://arxiv.org/abs/2211.00121} {arXiv:2211.00121 [cond-mat.str-el]}
  \BibitemShut {NoStop}%
\bibitem [{\citenamefont {Balodhi}\ and\ \citenamefont
  {Singh}(2017)}]{balodhi2017synthesis}%
  \BibitemOpen
  \bibfield  {author} {\bibinfo {author} {\bibfnamefont {A.}~\bibnamefont
  {Balodhi}}\ and\ \bibinfo {author} {\bibfnamefont {Y.}~\bibnamefont
  {Singh}},\ }\bibfield  {title} {\bibinfo {title} {Synthesis and pressure and
  field-dependent magnetic properties of the kagome-bilayer spin liquid ca 10
  cr 7 o 28},\ }\href@noop {} {\bibfield  {journal} {\bibinfo  {journal}
  {Physical Review Materials}\ }\textbf {\bibinfo {volume} {1}},\ \bibinfo
  {pages} {024407} (\bibinfo {year} {2017})}\BibitemShut {NoStop}%
\bibitem [{SM()}]{SM}%
  \BibitemOpen
  \href@noop {} {\bibinfo {title} {See {S}upplemental {M}aterial below for
  further information on the calculations reported throughout, which includes
  refs.~\cite{balz2017magnetic,wpd,mathematica,elstner1994spin,kshetrimayum2020tensor,balz2016physical}}}\BibitemShut
  {NoStop}%
\bibitem [{\citenamefont {Sugiura}\ and\ \citenamefont
  {Shimizu}(2013)}]{sugiura2013canonical}%
  \BibitemOpen
  \bibfield  {author} {\bibinfo {author} {\bibfnamefont {S.}~\bibnamefont
  {Sugiura}}\ and\ \bibinfo {author} {\bibfnamefont {A.}~\bibnamefont
  {Shimizu}},\ }\bibfield  {title} {\bibinfo {title} {Canonical {T}hermal
  {P}ure {Q}uantum {S}tate},\ }\href@noop {} {\bibfield  {journal} {\bibinfo
  {journal} {Phys. Rev. Lett.}\ }\textbf {\bibinfo {volume} {111}},\ \bibinfo
  {pages} {010401} (\bibinfo {year} {2013})}\BibitemShut {NoStop}%
\bibitem [{\citenamefont {Elstner}\ and\ \citenamefont
  {Young}(1994)}]{elstner1994spin}%
  \BibitemOpen
  \bibfield  {author} {\bibinfo {author} {\bibfnamefont {N.}~\bibnamefont
  {Elstner}}\ and\ \bibinfo {author} {\bibfnamefont {A.~P.}\ \bibnamefont
  {Young}},\ }\bibfield  {title} {\bibinfo {title} {Spin-1/2 {H}eisenberg
  antiferromagnet on the {\textit{kagom{\'e}}} lattice: {H}igh-temperature
  expansion and exact-diagonalization studies},\ }\href@noop {} {\bibfield
  {journal} {\bibinfo  {journal} {Phys. Rev. B}\ }\textbf {\bibinfo {volume}
  {50}},\ \bibinfo {pages} {6871} (\bibinfo {year} {1994})}\BibitemShut
  {NoStop}%
\bibitem [{\citenamefont {Schmidt}\ \emph {et~al.}(2011)\citenamefont
  {Schmidt}, \citenamefont {Lohmann},\ and\ \citenamefont
  {Richter}}]{schmidt2011eighth}%
  \BibitemOpen
  \bibfield  {author} {\bibinfo {author} {\bibfnamefont {H.-J.}\ \bibnamefont
  {Schmidt}}, \bibinfo {author} {\bibfnamefont {A.}~\bibnamefont {Lohmann}},\
  and\ \bibinfo {author} {\bibfnamefont {J.}~\bibnamefont {Richter}},\
  }\bibfield  {title} {\bibinfo {title} {Eighth-order high-temperature
  expansion for general {H}eisenberg {H}amiltonians},\ }\href@noop {}
  {\bibfield  {journal} {\bibinfo  {journal} {Phys. Rev. B}\ }\textbf {\bibinfo
  {volume} {84}},\ \bibinfo {pages} {104443} (\bibinfo {year}
  {2011})}\BibitemShut {NoStop}%
\bibitem [{\citenamefont {Lohmann}\ \emph {et~al.}(2014)\citenamefont
  {Lohmann}, \citenamefont {Schmidt},\ and\ \citenamefont
  {Richter}}]{lohmann2014tenth}%
  \BibitemOpen
  \bibfield  {author} {\bibinfo {author} {\bibfnamefont {A.}~\bibnamefont
  {Lohmann}}, \bibinfo {author} {\bibfnamefont {H.-J.}\ \bibnamefont
  {Schmidt}},\ and\ \bibinfo {author} {\bibfnamefont {J.}~\bibnamefont
  {Richter}},\ }\bibfield  {title} {\bibinfo {title} {Tenth-order
  high-temperature expansion for the susceptibility and the specific heat of
  spin-{$\bm{s}$} {H}eisenberg models with arbitrary exchange patterns:
  {A}pplication to pyrochlore and kagome magnets},\ }\href@noop {} {\bibfield
  {journal} {\bibinfo  {journal} {Phys. Rev. B}\ }\textbf {\bibinfo {volume}
  {89}},\ \bibinfo {pages} {014415} (\bibinfo {year} {2014})}\BibitemShut
  {NoStop}%
\bibitem [{\citenamefont {Blundell}(2001)}]{blundell2001magnetism}%
  \BibitemOpen
  \bibfield  {author} {\bibinfo {author} {\bibfnamefont {S.}~\bibnamefont
  {Blundell}},\ }\href@noop {} {\emph {\bibinfo {title} {Magnetism in
  {C}ondensed {M}atter}}}\ (\bibinfo  {publisher} {OUP Oxford},\ \bibinfo
  {year} {2001})\BibitemShut {NoStop}%
\bibitem [{inp()}]{inprep1}%
  \BibitemOpen
  \href@noop {} {\bibinfo {title} {{J. A. Crossley and C. A. Hooley, in
  preparation.}}}\BibitemShut {Stop}%
\bibitem [{wpd()}]{wpd}%
  \BibitemOpen
  \href@noop {} {}\bibinfo {howpublished}
  {\url{https://automeris.io/WebPlotDigitizer/}}\BibitemShut {NoStop}%
\bibitem [{mat()}]{mathematica}%
  \BibitemOpen
  \href@noop {} {}\bibinfo {howpublished}
  {\url{https://www.wolfram.com/mathematica/}}\BibitemShut {NoStop}%
\end{thebibliography}%

\onecolumngrid

\newpage

\begin{center}
\textbf{\large Supplemental Material for ``What is the origin of the intermediate-temperature magnetic specific heat capacity in the spin-liquid candidate \cro?"}
\end{center}

\setcounter{section}{0}
\setcounter{figure}{0}
\renewcommand{\thefigure}{S\arabic{figure}}

\vspace*{20mm}
\section{Defining a unit cell}
In order to make direct comparison with the published experimental specific heat capacity of Balz~\textit{et~al.}~we had to adopt the same definition of the unit cell.

To do this, we referred to the magnetic specific heat capacity reported in Fig.~2\,(c) of ref.\,\cite{balz2017magnetic}. Integrating $C/T$ between $T = 0$ and $T \rightarrow \infty$, one would obtain the change in molar magnetic entropy in bringing the system out of its ground state manifold towards the high-temperature limit where every microstate is equally accessible. We focused on the $12\,\rm{T}$ case where there is a unique, fully-polarized ground state. For $12\,\rm{T}$ therefore, the \textit{change} in molar entropy (in raising the temperature from $0\,\rm{K}$ to the infinite limit) is equal to the \textit{total} molar entropy of the system.

Using WebPlotDigitizer \cite{wpd}, we extracted the high-field data from Fig.~2\,(c) of ref.\,\cite{balz2017magnetic}. Using Mathematica \cite{mathematica}, we converted the data to $C/T$, created a continuous curve using the `Interpolate' function, and then performed numerical integration and simultaneous extrapolation up to $23\,\rm{K}$ (in order to approximately capture as much of the available entropy of the system as possible) using `NIntegrate'. From this we deduced a total entropy change of ${\sim}\,33.5\,\mathrm{J}\,\mathrm{mol}^{-1}\mathrm{K}^{-1}$. We expect the exact answer to be equal to $nR\,\mathrm{ln}(2)$, where $R\, \mathrm{ln}(2)$ is the maximum available entropy per mole of a given spin--$\frac{1}{2}$ site and $n$ is the number of sites per unit cell. Dividing 33.5 by $R\, \mathrm{ln}(2)$ we get ${\sim}\,5.8$ from which we may assume that $n = 6$ (6 sites per unit cell). This coincides with the size of the unit cell for a single kagome bilayer of $\mathrm{Cr}^{5+}$ ions as shown in Fig. 1 of the main article.
\vspace{20mm}
\section{Calculation details}
\subsection{Coupling parameters}
Table\ \ref{couplings} displays the different coupling values we used across this work. For the strong-triangle method we used column (a), for exact diagonalization we used column (b), for Thermal Pure Quantum States and the high-temperature expansion up to $5^{\mathrm{th}}$ order we used column (c), and for all other high-temperature expansion calculations we used column (d). Cases where we have set couplings to zero are clearly signposted in the main text. Otherwise, all coupling values used fall within the uncertainty margins for the couplings of the BBK model for \cro\ given by Balz~\textit{et~al}. \cite{balz2017magnetic}.
\begin{table*}[h!]
\centering
\begin{tabular}{ccccc}
 Coupling / meV&(a)&(b)&(c)&(d)\\ \hline
 $J_{\Delta1}$&$-0.2730$&$-0.2730$&$-0.2730$&$-0.2730$\\
 $J_{\nabla1}$&$0$&$0.0920$&$0.0920$&$0.1025$\\
 $J_{\Delta2}$&$0$&$0.1130$&$0.1130$&$0.1025$\\
$J_{\nabla2}$&$-0.7650$&$-0.7650$&$-0.7650$&$-0.7650$\\
 $J_{\rm inter}$&$0$&$0$&$-0.0840$&$-0.0840$\\
\end{tabular}
\caption{\label{couplings}The four different sets of coupling values used in the calculations throughout this work. Couplings are stated in units of meV.}
\end{table*}

\subsection{Cluster definitions}
\begin{figure}[h!]
    \centering
    \subfloat[]{\includegraphics[trim=50ex 0ex 50ex 0ex, clip, width=0.45\textwidth]{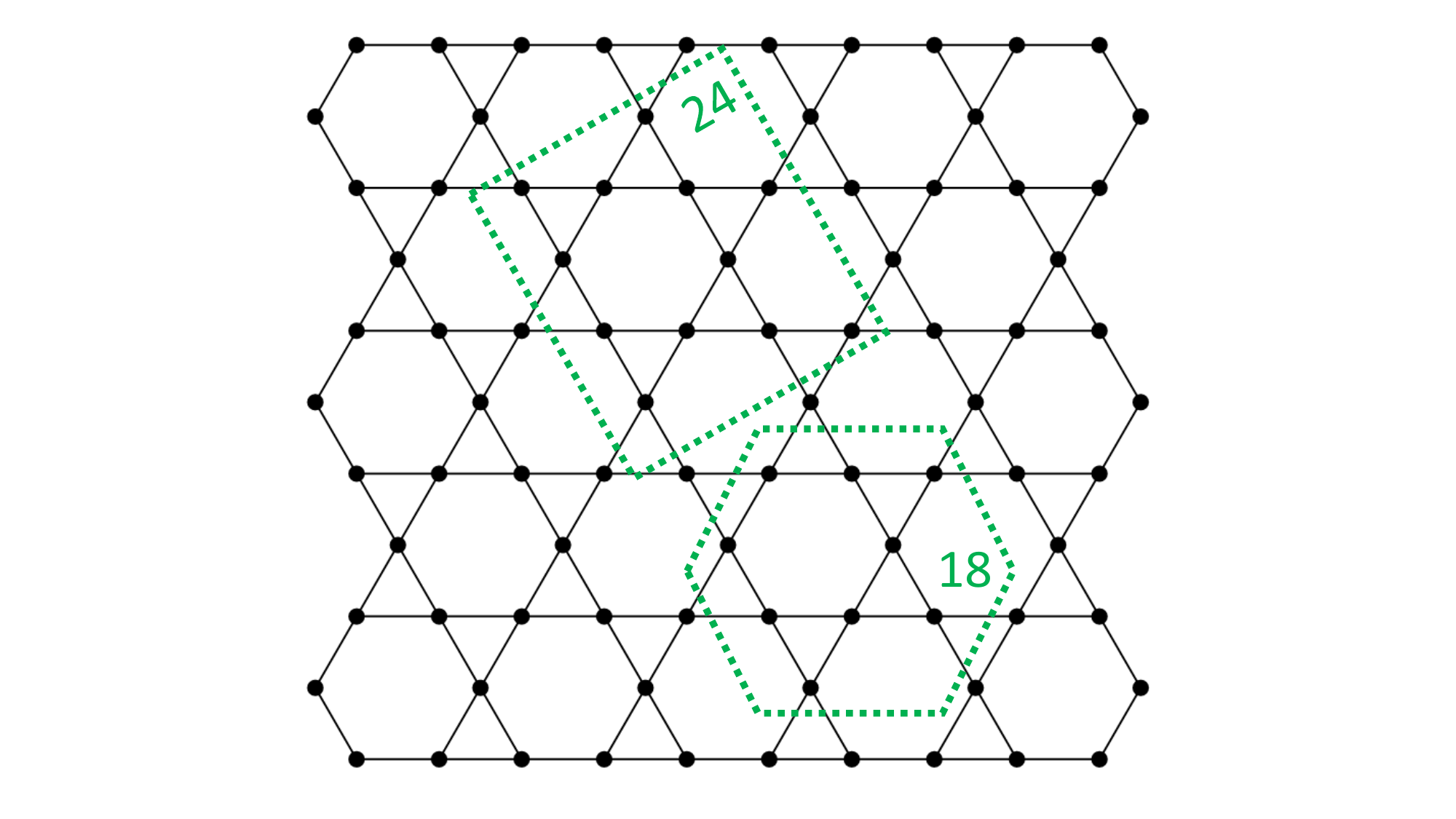}\label{smaller}}\hskip5ex
    \subfloat[]{\includegraphics[trim=50ex 0ex 50ex 0ex, clip, width=0.45\textwidth]{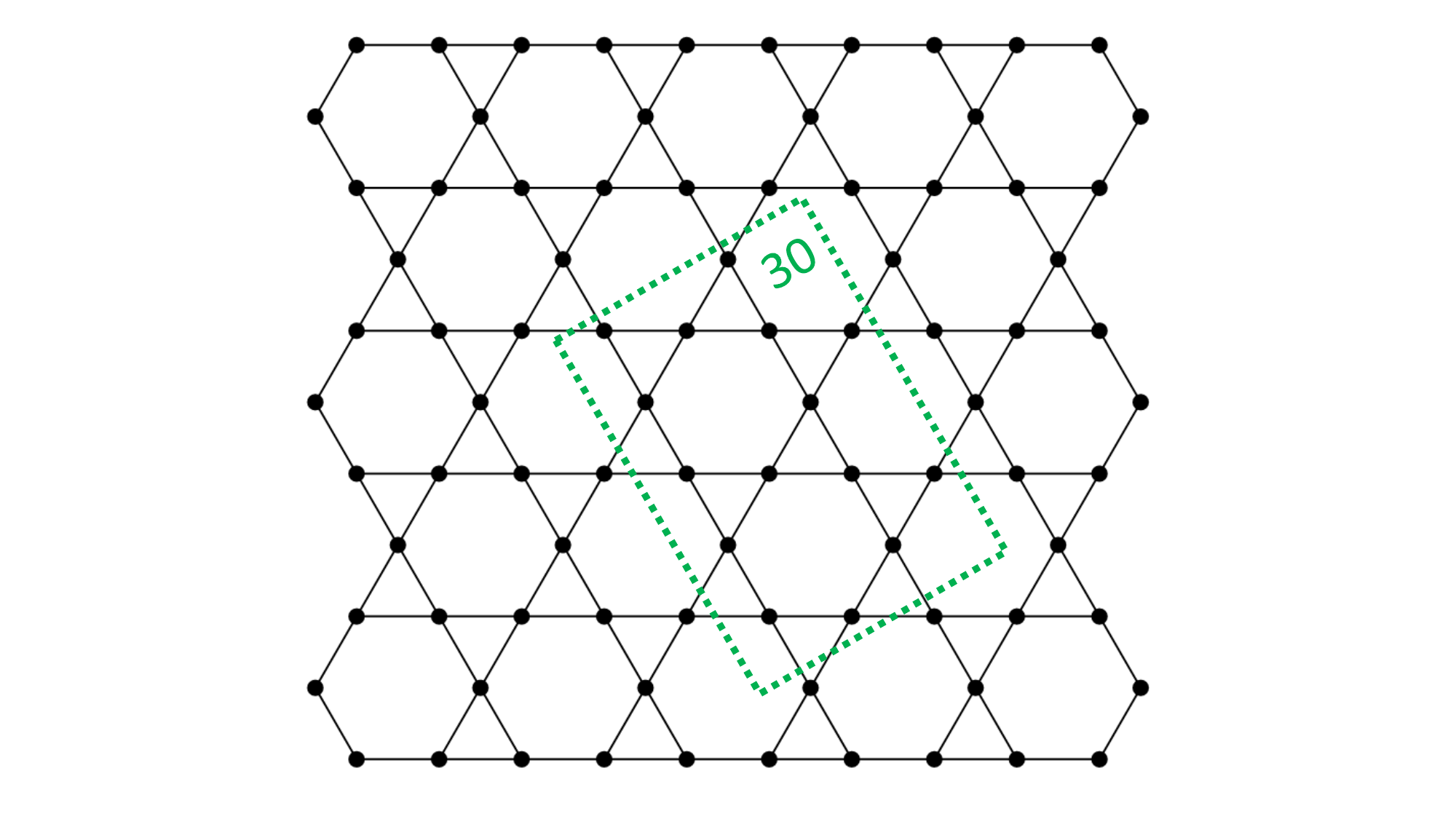}\label{larger}}
\caption{Diagrams of the 18, 24, and 30-spin clusters used in all exact diagonalization and Thermal Pure Quantum States calculations throughout this paper. For clarity, only a single layer of the bilayer structure is shown; one should imagine each dotted boundary enclosing the same set of spins in the other layer also. The 24-spin cluster can tile space in two different ways but the resulting Hamiltonian is the same in both cases.}
\label{clusters}
\end{figure}

We have reported a range of calculations performed on finite clusters of spins with periodic boundary conditions. Fig.~\ref{clusters}\,(a) shows the clusters used in all calculations involving either 18 or 24 spins. Fig.~\ref{clusters}\,(b) shows the cluster used in all 30-spin calculations.

\begin{figure}[h!]
    \centering
    \subfloat[]{\includegraphics[width=0.3\textwidth]{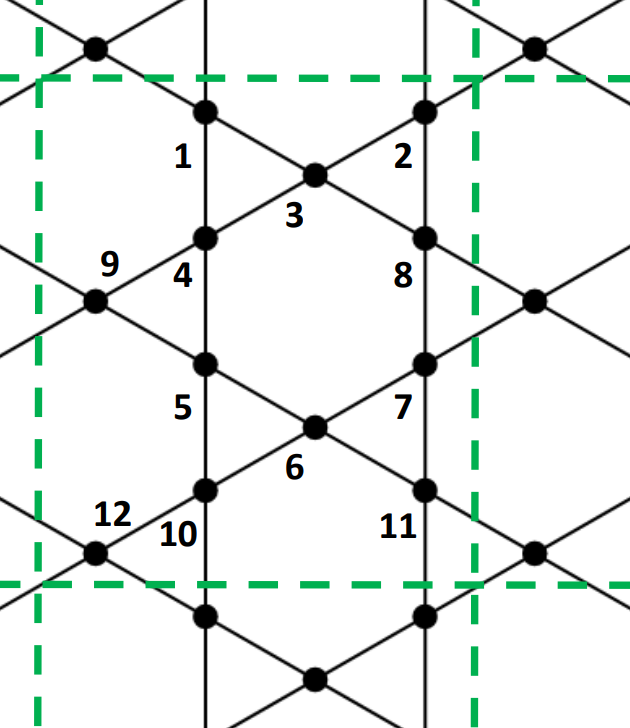}\label{24_aligned}}\hskip25ex
    \subfloat[]{\includegraphics[width=0.3\textwidth]{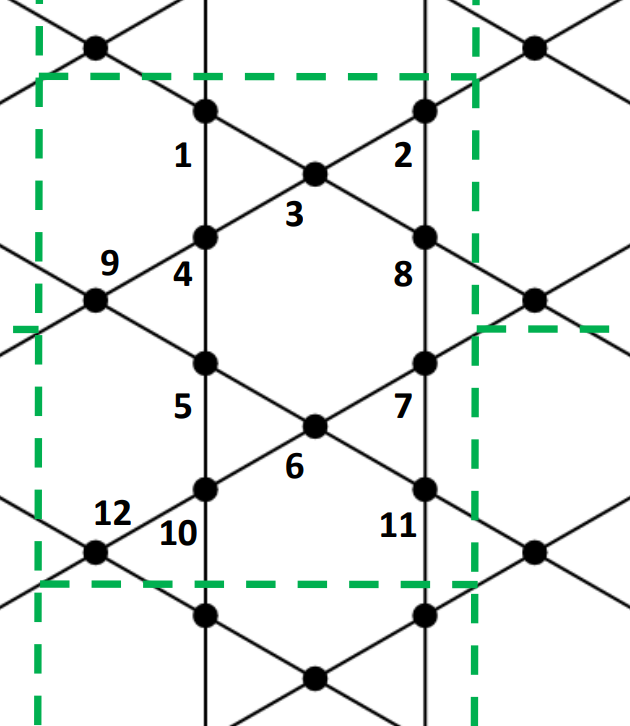}\label{24_shifted}}
\caption{The two alternative tilings of the 24-spin cluster outlined in Fig.~\ref{clusters}\,(a). We show a single layer of the bilayer structure for clarity, and we ascribe labels to the depicted spins in correspondence with the labels used in Fig.~\ref{graphs}.}
\label{labelled}
\end{figure}

The 24-spin cluster can be used to tile the lattice in two different ways: checkered tiling (Fig.~\ref{labelled}\,(a)) and brick-wall tiling (Fig.~\ref{labelled}\,(b)).

\begin{figure}[h!]
    \centering
    \subfloat[]{\includegraphics[width=0.4\textwidth]{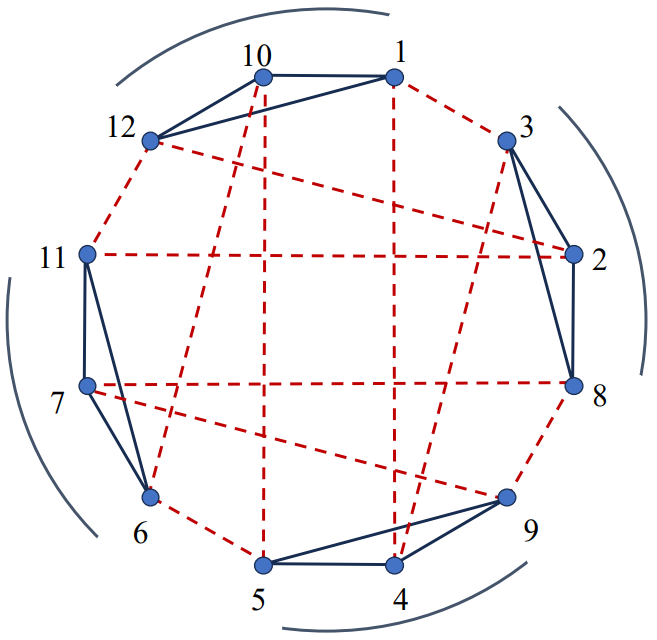}\label{graph_aligned}}\hskip14ex
    \subfloat[]{\includegraphics[width=0.4\textwidth]{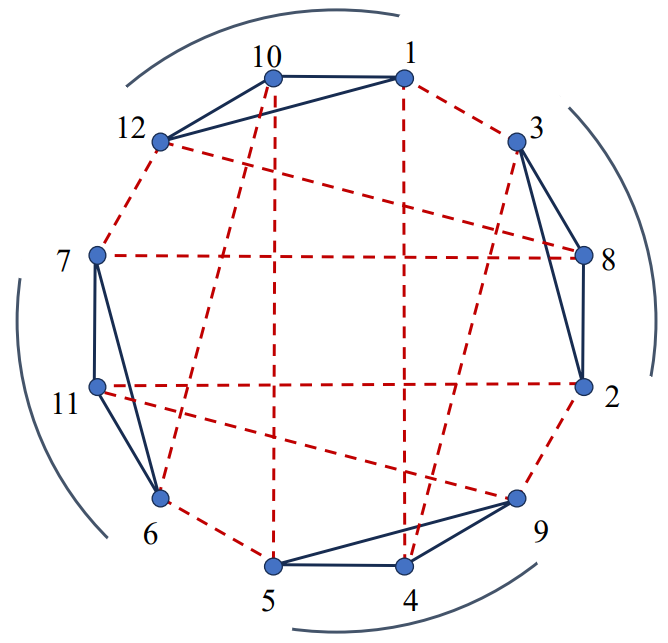}\label{graph_shifted}}
\caption{Graphs showing the connectivity of a single layer of spins for the two alternative tilings of the 24-spin cluster outlined in Fig.~\ref{clusters}\,(a). Graph (a) corresponds to the checkered tiling of Fig.~\ref{labelled}\,(a), and graph (b) corresponds to the brick-wall tiling of Fig.~\ref{labelled}\,(b). Red hatched lines connecting vertices correspond to bonds within the right-pointing triangles of Fig.~\ref{labelled}, while black lines connecting vertices correspond to bonds within the left-pointing triangles. The arcs surrounding each graph have been added to help elucidate the $C_{4}$ symmetry.}
\label{graphs}
\end{figure}

These two tilings simply differ by site labelling. To demonstrate this, Fig.~\ref{graphs} shows their corresponding connectivity.
\section{Benchmarking the TPQS code}
\begin{figure}[h!]
    \centering
    \includegraphics[scale=0.8,trim=8ex 0ex 0ex 3ex, clip]{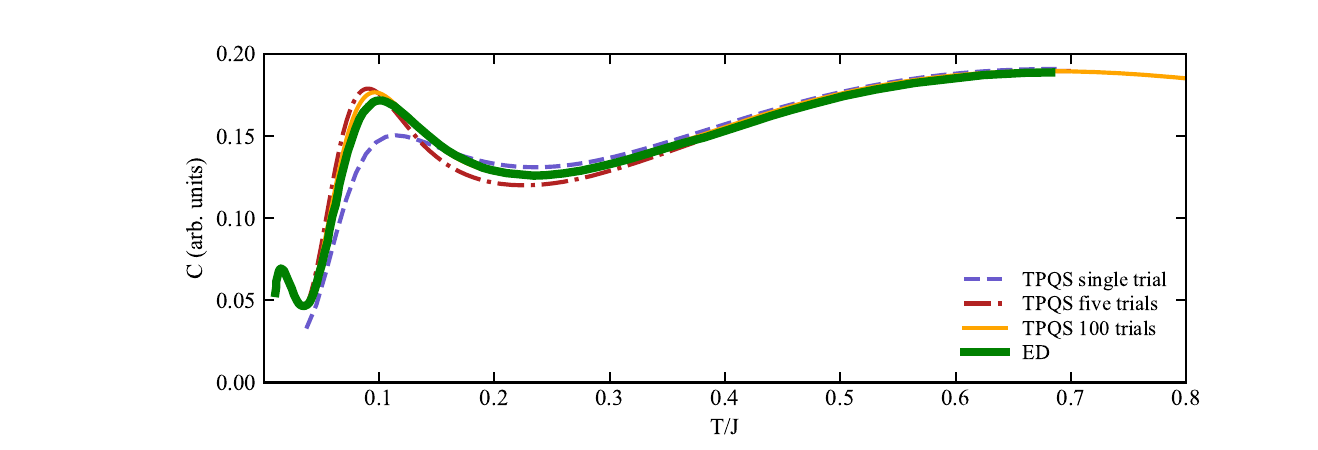}
    \vspace{-5mm}
    \caption{The zero-field magnetic specific heat capacity of a cluster of 18 spin--$\frac{1}{2}$ moments under nearest-neighbor antiferromagnetic Heisenberg interactions on the kagome lattice. We show TPQS results for a single trial (blue dashed curve), a five-trial average (red dot-dashed curve), and a 100-trial average (yellow curve). The exact diagonalization data (green) of Elstner and Young \cite{elstner1994spin} was copied using WebPlotDigitizer \cite{wpd} and is shown here for comparison.
    \label{Kag}}
\end{figure}

To benchmark our TPQS code we calculated the specific heat capacity for a cluster of 18 spin--$\frac{1}{2}$ moments on a kagome lattice interacting under nearest-neighbor antiferromagnetic Heisenberg interactions. This calculation has already been done exactly by Elstner and Young \cite{elstner1994spin}, enabling a direct comparison to be made. The uncertainties associated with using this technique for such a small cluster of spins are relatively high and so we chose to loop the algorithm 100 times, taking the average result. This lowered the uncertainty bounds on our results by a factor of ten. Fig.~\ref{Kag} demonstrates our results converging towards the exact diagonalization data as we average over successively more TPQS trials.
\newpage
\section{An entropic comparison of theory and experiment}
\begin{figure}[h!]
    \centering
    \includegraphics[trim=0ex 0ex 0ex 18ex, clip, width=0.85\textwidth]{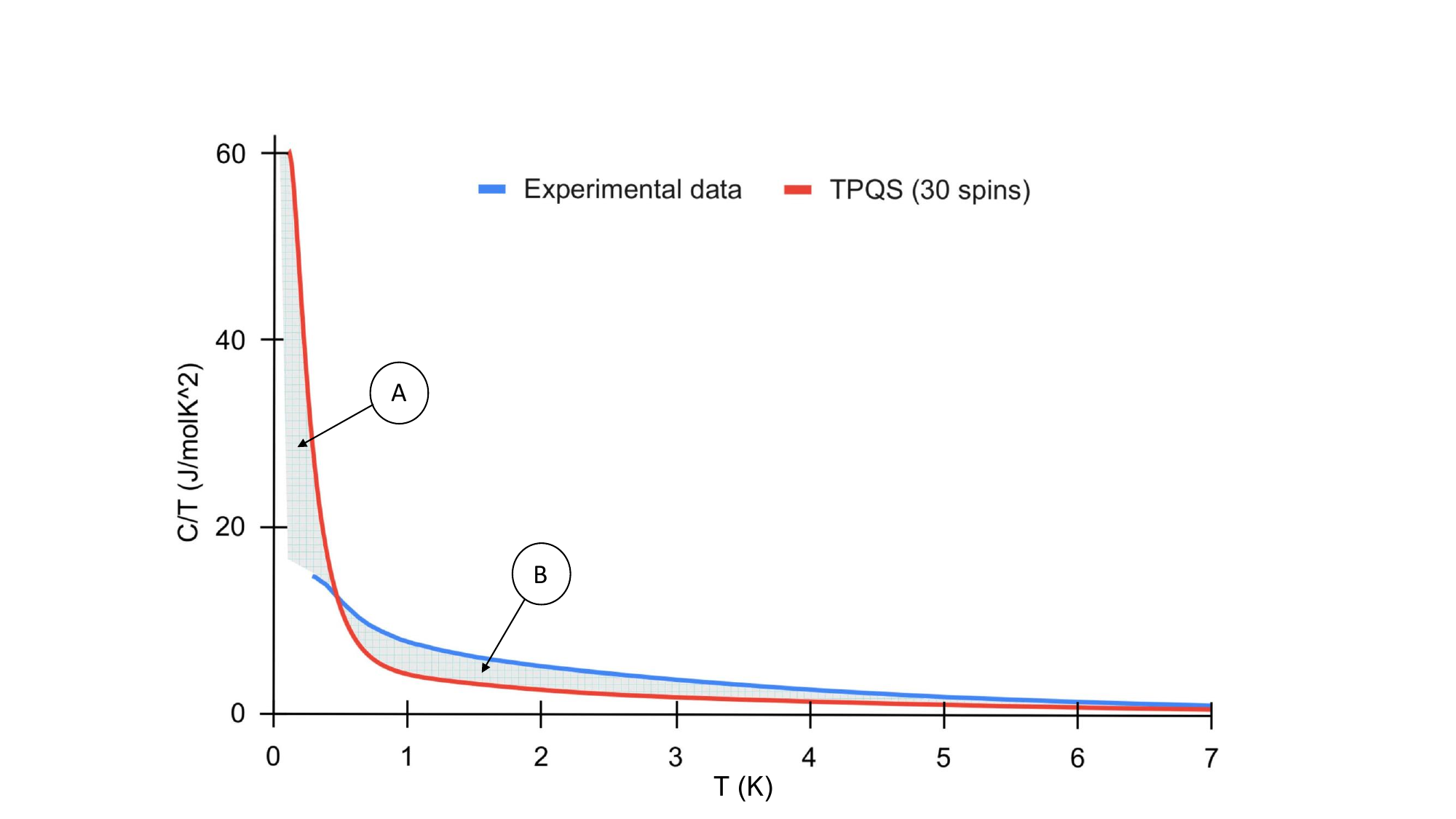}
    \caption{A plot of the zero-field magnetic specific heat capacity divided by temperature for \cro. The experimental curve (blue) was obtained by linear interpolation of data points copied from ref.\,\cite{kshetrimayum2020tensor} using WebPlotDigitizer \cite{wpd}. The two curves cross over at around $T = 0.45\,\rm{K}$. We note that the entropy surplus in the TPQS simulation relative to the experiment at low temperatures (\textit{A}) appears to balance the entropy deficit at higher temperatures (\textit{B}).}
    \label{C/T}
\end{figure}

In Fig.~\ref{C/T} we compare the $C/T$ curve for our 30-spin TPQS calculation with that of experiment. The experimental curve was obtained by linear interpolation of data points copied from ref.\,\cite{kshetrimayum2020tensor} using WebPlotDigitizer \cite{wpd}. The shaded areas labelled \textit{A} and \textit{B} correspond to differences in the entropy under each curve. Area \textit{A} approximately corresponds to the surplus molar entropy in the 30-spin TPQS simulation (relative to the experiment) upon raising the temperature from $0.15\,\mathrm{K}$ to $0.45\,\mathrm{K}$. Area \textit{B} corresponds to the extra molar entropy unlocked in experiment (relative to the 30-spin TPQS simulation) upon raising the temperature from $0.45\,\mathrm{K}$ to $ 7\,\mathrm{K}$. The two curves essentially converge beyond this point. These two areas appear roughly equal, and therefore the curves contain similar entropies as one would hope. However, this argument is hardly rigorous; we do not have sufficient data to see exactly how the two curves would bound the lower-temperature part of the area \textit{A}.
\vspace{20mm}
\section{The specific heat capacity of the coupled hexagon model}
\begin{figure}[h!]
    \centering
    \includegraphics[scale=0.75,trim={1cm 0 0 0}]{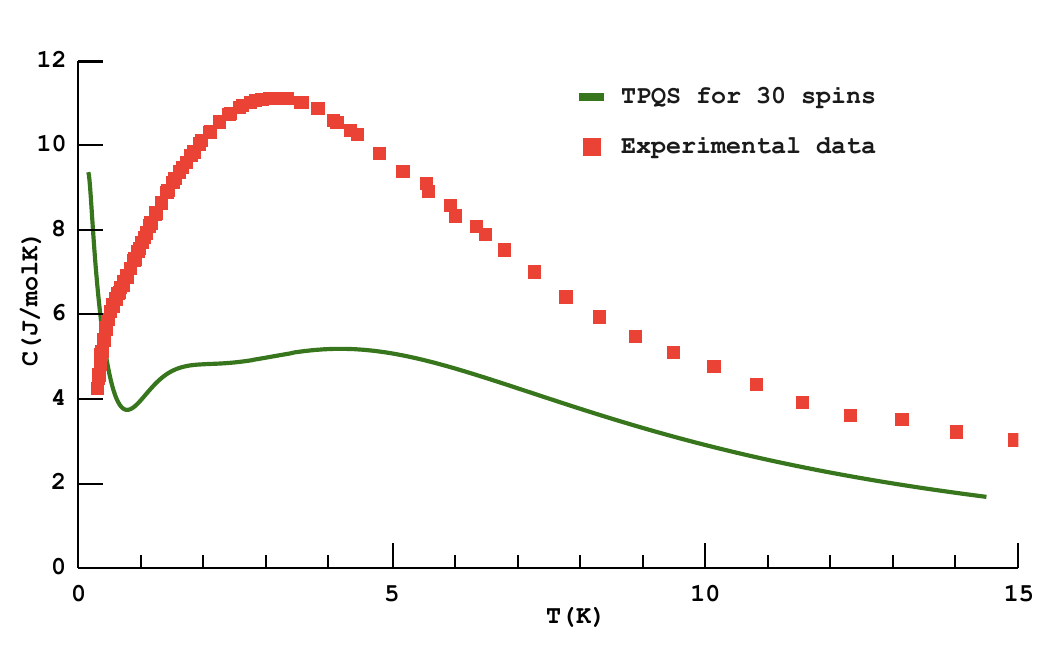}
    \caption{The zero-field magnetic specific heat capacity of the coupled hexagon model for \cro. We performed the calculation for a cluster of 30 spins using the Thermal Pure Quantum States algorithm (green curve). This model was first presented by Balz~\textit{et~al.}~in ref.\,\cite{balz2017magnetic}. The experimental zero-field specific heat capacity of \cro, previously published in ref.\,\cite{balz2017magnetic}, is given for reference (red squares).}
    \label{alt}
\end{figure}

Throughout this report we have considered the breathing bilayer kagome (BBK) model for \cro\ as presented by Balz~\textit{et~al.}~in ref.\,\cite{balz2016physical}. The same authors mention an alternative magnetic Hamiltonian for this material, called the coupled hexagon model \cite{balz2017magnetic}. These two models fit the high-field neutron scattering data for the real material equally well, yet the BBK model is favored on the basis of structural considerations. We chose to revisit the coupled hexagon model by calculating the specific heat capacity of a cluster of 30 spins using the TPQS algorithm. Our results are shown in Fig.~\ref{alt}. They closely resemble those of the BBK model: a strong peak below $1\,\rm{K}$, and a diminished broad peak around $3-5\,\rm{K}$. Hence we see that the predictions of the coupled hexagon model deviate from the experimental data in much the same way as those of the BBK model, and thus the coupled hexagon model does no better than the BBK model at reproducing the experimental specific heat capacity for a small cluster of spins.
\vspace*{18mm}
\section{Varying the BBK model coupling constants}
To investigate whether the discrepancy between theory and experiment could be remedied simply by an alteration of the coupling constants, we identified three close relatives of the BBK model (\ref{BBK}) whose 0 T specific heat capacities match that of \cro\ at temperatures that we can easily simulate. These alternative models have the same arrangement of couplings as the BBK model (Fig.~\ref{structure}), but their exchange strengths differ in various ways, as shown in Table\ \ref{relatives}.
\begin{table*}[h!]
\centering
\setlength\tabcolsep{2pt}
\begin{tabular}{c|ccccc}
 Coupling / meV&$J_{\Delta1}$&$J_{\nabla1}$&$J_{\Delta2}$&$J_{\nabla2}$&$J_{\rm inter}$\\ \hline
 original&$-0.27$&$0.09$&$0.11$&$-0.76$&$-0.08$\\
 alt-1&$-0.273$&$0.1025$&$0.1025$&$-0.765$&$0.6$\\
\end{tabular}
\quad
\begin{tabular}{c|ccccc}
 Coupling / meV&$J_{\Delta1}$&$J_{\nabla1}$&$J_{\Delta2}$&$J_{\nabla2}$&$J_{\rm inter}$\\ \hline
 alt-2&$-0.7$&$0.1025$&$0.1025$&$-0.76$&$0.01$\\
 alt-3&$-0.5$&$0.5$&$0.5$&$-0.5$&$-0.1$\\
\end{tabular}
\caption{\label{relatives}The exchange-coupling strengths of the original breathing bilayer kagome (BBK) model, followed by three closely related alternatives labelled alt-\{1,2,3\}. Couplings are stated in units of meV.}
\end{table*}

These alternative sets of exchange couplings have not been determined by any microscopic calculation but simply chosen to fit the high-temperature tail of the 0 T specific heat capacity for \cro. The question is: do they still fit the 11 T single-spin-flip dispersion relations? In Fig.~\ref{bad_relatives} we show a comparison between the single-spin-flip dispersion relations predicted by these alternative models and the experimental data from Fig.~12 of ref.\,\cite{balz2017magnetic}. One would find it very hard to argue that the calculated dispersion relations for any of these models are consistent with the bands seen in the 11 T inelastic neutron scattering data for \cro. We therefore reject all three of these alternative models, and tentatively conclude that there is unlikely to be a purely Heisenberg model that fits all of the currently available data on \cro.
\newpage
\newcommand{\scale}{0.33}
\newcommand{\scalelong}{0.57}
\begin{figure}[h!]
    \centering
    \includegraphics[scale=\scale]{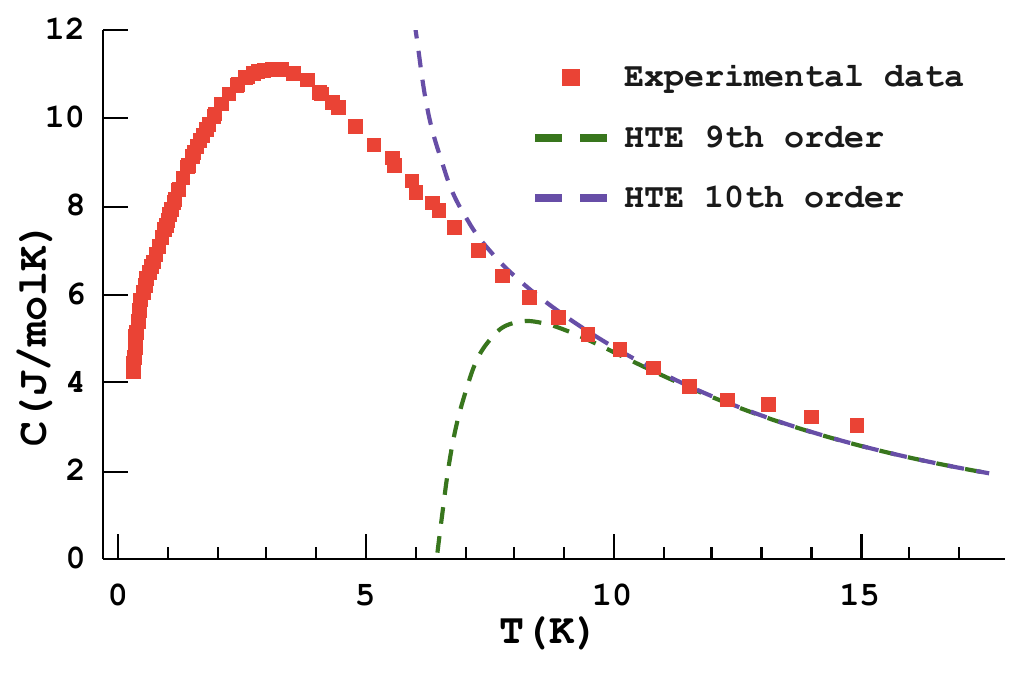}\hspace*{-3mm}
    \includegraphics[scale=\scalelong]{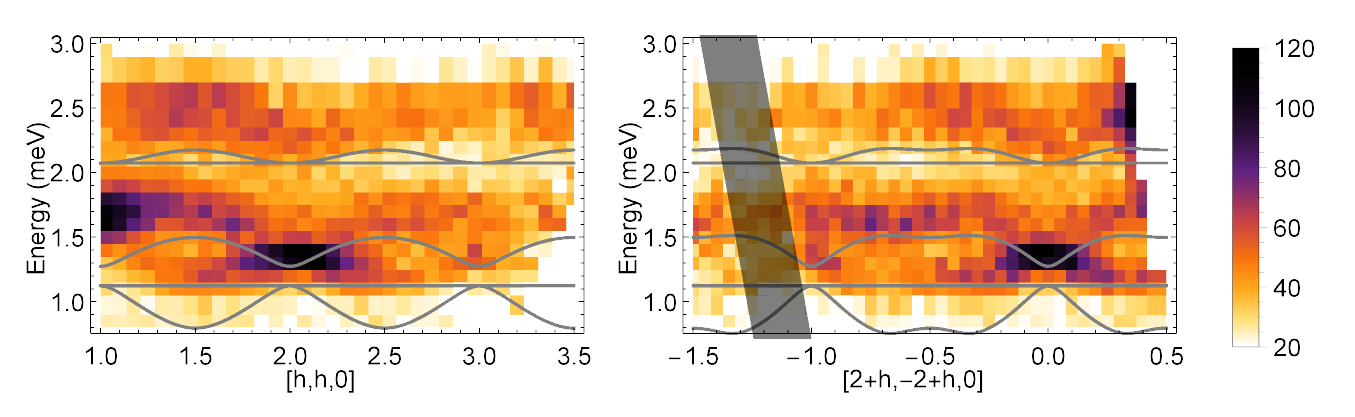}\\
    \vspace*{-4mm}
    \includegraphics[scale=\scale]{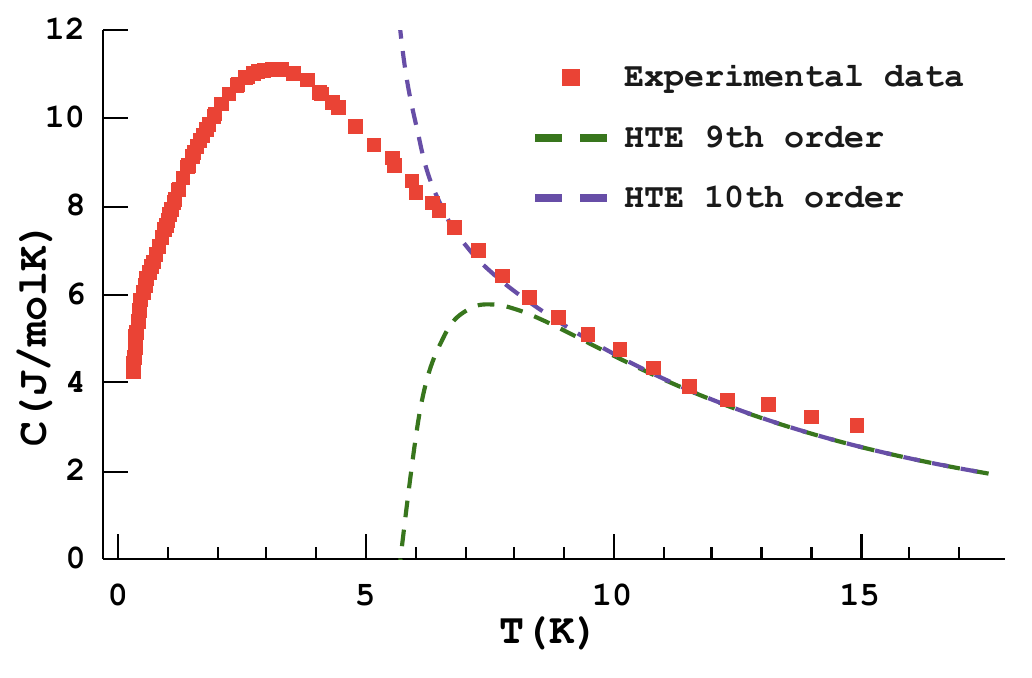}\hspace*{-3mm}
    \includegraphics[scale=\scalelong]{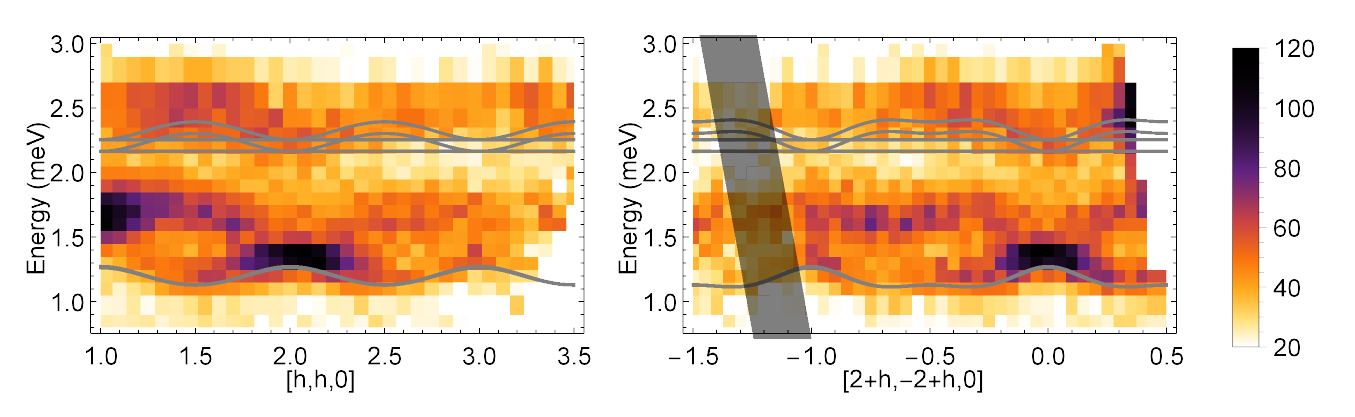}\\
    \vspace*{-4mm}
   \includegraphics[scale=\scale]{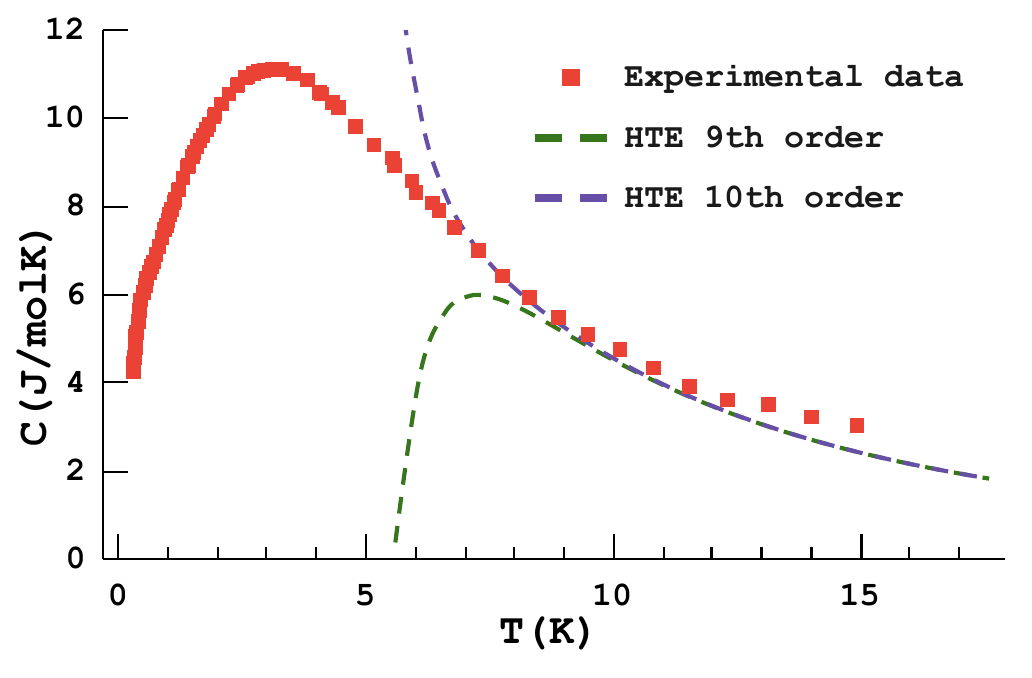}\hspace*{-3mm}
    \includegraphics[scale=\scalelong]{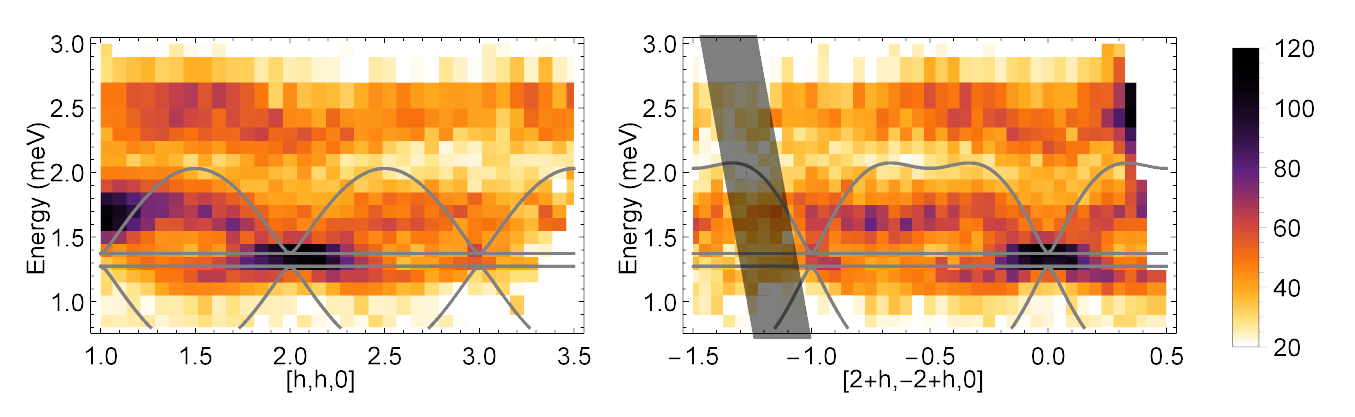}\\
\caption{The specific heat capacity (dashed lines, left) and single-spin-flip dispersion bands (gray lines, center \& right) for three alternatives to the breathing bilayer kagome (BBK) model. From top to bottom, the rows correspond to models alt-1, alt-2, and alt-3, definitions for which can be found in Table\ \ref{relatives}. The specific heat capacity was calculated using the high-temperature expansion. The experimental specific heat capacity and 11 T inelastic neutron scattering (INS) data for \cro\ are given for comparison \cite{balz2017magnetic}. The shaded region of the INS plots shows the area of artificially low intensity as per ref.\,\cite{balz2017magnetic}; diagrams of the relevant momentum-space cuts can be found in ref.\,\cite{balz2017magnetic}, too.}
\label{bad_relatives}
\end{figure}
\end{document}